\documentclass[%
reprint,
superscriptaddress,
showpacs,preprintnumbers,
amsmath,amssymb,
prb,
]{revtex4-1}

\usepackage{graphicx}
\usepackage{dcolumn}
\usepackage{bm}
\usepackage{xcolor}

\begin{document}

\newcommand{\Rb}{RbFe$_2$As$_2$}
\newcommand{\Cs}{CsFe$_2$As$_2$}
\newcommand{\K}{KFe$_2$As$_2$}
\newcommand{\Tc}{$T_{\rm c}$}
\newcommand{\Tn}{$T_{\rm N}$}
\newcommand{\Pc}{$P_{\rm c}$}
\newcommand{\PH}{$P_{\rm H}$}
\newcommand{\dhdt}{$\left(-\partial{\rm H_{c2}}/\partial{\rm T} \right)_{T_{c\rm}}$}
\newcommand{\ie}{{\it i.e.}}
\newcommand{\eg}{{\it e.g.}}
\newcommand{\etal}{{\it et al.}}  
\newcommand{\mucm}{$\mathrm{\mu\Omega\, cm}$}
\newcommand{\Tmin}{$T_{\rm min}$}
\newcommand{\Tstar}{$T^\star$}
\newcommand{\Hvs}{$H_{\rm vs}$}
\newcommand{\Hc}{$H_{\rm c2}$}
\newcommand{\Hcstar}{$H_{\rm c2}^\star$}
\newcommand{\Hstar}{$H^\star$}
\newcommand{\Nqp}{$N_{\rm qp}$}
\newcommand{\Nsc}{$N_{\rm sc}$}
\newcommand{\NbSi}{Nb$_{0.15}$Si$_{0.85}$}
\newcommand{\RH}{$R_{\rm H}$}
\newcommand{\xc}{$x_{\rm c}$}
\newcommand{\LB}{$\ell_{\rm B}$}


\title{Tuning the electronic and the crystalline structure of LaBi by pressure}

\author{F. F. Tafti}
\email{fazel.tafti@bc.edu}
\affiliation{Department of Physics, Boston College, Boston, MA, USA}
\affiliation{Department of Chemistry, Princeton University, NJ, USA}

\author{M. S. Torikachvili}
\affiliation{Department of Physics, San Diego State University, San Diego, CA, USA}

\author{R. L. Stillwell}
\affiliation{Lawrence Livermore National Laboratory, Livermore, CA, USA}

\author{B. Baer}
\affiliation{Lawrence Livermore National Laboratory, Livermore, CA, USA}

\author{E. Stavrou}
\affiliation{Lawrence Livermore National Laboratory, Livermore, CA, USA}

\author{S. T. Weir}
\affiliation{Lawrence Livermore National Laboratory, Livermore, CA, USA}

\author{Y. K. Vohra}
\affiliation{Department of Physics, University of Alabama at Birmingham, Birmingham, AB, USA}

\author{H.-Y. Yang}
\affiliation{Department of Physics, Boston College, Boston, MA, USA}

\author{E. F. McDonnell}
\affiliation{Department of Physics, Boston College, Boston, MA, USA}

\author{S. K. Kushwaha}
\affiliation{Department of Chemistry, Princeton University, NJ, USA}

\author{Q. D. Gibson}
\affiliation{Department of Chemistry, Princeton University, NJ, USA}

\author{R. J. Cava}
\affiliation{Department of Chemistry, Princeton University, NJ, USA}

\author{J. R. Jeffries}
\affiliation{Lawrence Livermore National Laboratory, Livermore, CA, USA}

\date{\today}

\begin{abstract}
Extreme magnetoresistance (XMR) in topological semimetals is a recent discovery which attracts attention due to its robust appearance in a growing number of materials.
%
%
To search for a relation between XMR and superconductivity, we study the effect of pressure on LaBi taking advantage of its simple structure and simple composition.
By increasing pressure we observe the disappearance of XMR followed by the appearance of superconductivity at $P\approx 3.5$ GPa.
The suppression of XMR is correlated with increasing zero-field resistance instead of decreasing in-field resistance.
At higher pressures, $P\approx 11$ GPa, we find a structural transition from the face center cubic lattice to a primitive tetragonal lattice in agreement with theoretical predictions.
We discuss the relationship between extreme magnetoresistance, superconductivity, and structural transition in LaBi.

\end{abstract}

\pacs{73.43.Qt, 64.70.K-, 74.62.Fj, 71.20.Lp}
\maketitle



\section{\label{Introduction}Introduction}

Extreme magnetoresistance is an enormous increase of electrical resistance in response to a modest magnetic field in several topological semimetals including Cd$_3$As$_2$, Na$_3$Bi, NbAs, NbP, TaAs, NbSb$_2$, TaSb$_2$, WTe$_2$, (Zr/Hf)Te$_5$. \cite{liang_ultrahigh_2015, xiong_evidence_2015, shekhar_extremely_2015, ghimire_magnetotransport_2015, huang_observation_2015, wang_anisotropic_2014, wang_topological_2016, ali_large_2014, ali_correlation_2015, tritt_large_1999}
Recent studies on (W/Mo)Te$_2$ and (Zr/Hf)Te$_5$ suggest that pressure suppresses the extreme magnetoresistance (XMR) and gives rise to superconductivity. \cite{kang_superconductivity_2015, pan_pressure-driven_2015, zhou_pressure-induced_2016, qi_pressure-driven_2016} 
Common to all these materials is a rapid onset of superconductivity at the pressure where XMR is suppressed, followed by a slow suppression of $T_c$ with further increasing pressure.
For example, MoTe$_2$ is superconducting at zero pressure with $T_c=0.1$ K which rapidly increases to 8 K by applying only 1 GPa of pressure. \cite{qi_superconductivity_2016}
WTe$_2$ is not superconducting at $P=0$, it shows an incomplete superconducting transition at $P=2.5$ GPa and a full transition at $P=8$ GPa. \cite{kang_superconductivity_2015, pan_pressure-driven_2015}
Similarly, ZrTe$_5$ is not superconducting at $P=0$, it shows a sudden onset of superconductivity at $P=6.7$ GPa with a subsequent $T_c$ discontinuity at $P=20$ GPa attributed to a second superconducting state. \cite{zhou_pressure-induced_2016}
By pressurizing LaBi we reveal all the above-mentioned characteristics including XMR suppression, superconducting transition, and discontinuous $T_c$ evolution in a single material.

In this work, we confirm the link between XMR and superconductivity by studying LaBi, a recent topological semimetal that attracted attention due to its simple lattice and band structures. \cite{zeng_topological_2015, tafti_temperaturefield_2016, tafti_resistivity_2016, zeng_compensated_2016, wu_unusual_2016, nayak_multiple_2016}
The three panels of Fig. \ref{PD} summarize our main findings:
(a) The suppression of XMR by pressure is a purely electronic effect with no drastic changes in structural parameters, 
(b) Pressure suppresses XMR and induces superconductivity, and
(c) The discontinuity in $T_c$ at higher pressures is due to a structural transition.



\section{\label{Experiments}Methods}

Single crystals of LaBi were grown using indium flux and characterized using powder x-ray diffraction and energy dispersive x-ray spectroscopy as explained in previous works. \cite{tafti_temperaturefield_2016, sun_large_2016}
Low pressure measurements ($P<2$ GPa) were performed in a piston-cylinder clamp cell using 40:60 mixture of light mineral oil:n-pentane as a hydrostatic medium.
Pressure was measured from the superconducting transition of a Pb gauge placed beside the sample in the clamp cell. \cite{eiling_pressure_1981}
The pressure cell was fit to a Quantum Design PPMS which monitored simultaneously the resistance of the sample, the Pb gauge, and a calibrated cernox sensor attached to the body of the cell for accurate thermometry. 
High pressure measurements were performed in a designer diamond anvil cell using steatite as the pressure transmitting medium and MP35N as the gasket material. \cite{weir_epitaxial_2000}  
The designer diamond had eight tungsten micro-contacts centered on a 300 $\mu$m culet for electrical transport measurements.
Pressure was measured by fluorescent spectroscopy on two pieces of ruby placed beside the sample in the diamond anvil cell. \cite{piermarini_calibration_1975}  
A small single crystal of LaBi ($50\times 50 \times 10$ $\mu$m) was placed inside the 120 $\mu$m diameter sample hole made by the electric discharge method.
%
%
High pressure x-ray diffraction was performed in a membrane driven DAC with 300 $\mu$m culet diamond anvils and rhenium gasket with 120 $\mu$m hole filled with LaBi powder, copper powder as the pressure marker, and neon as the hydrostatic medium. 
Diffraction experiments took place at the Advanced Photon Source at Argonne National Laboratory (beamlines 16 ID-B and 13 ID-D) with 29.2 and 37.1 keV monochromatic x-ray beam. 
Angle dispersive diffraction patterns were collected with an area detector (Pilatus1M or Mar345) with exposure times ranging from 20-120 seconds. 
Two-dimensional x-ray diffraction images were integrated using FIT2D \cite{hammersley_two-dimensional_1996} software and refined using the EXPGUI/GSAS \cite{toby_xpgui_2001} software to extract structural parameters.
Band structure calculations are performed with the WIEN2k program using the general gradient approximation on augmented plane-waves and local orbitals. \cite{blaha_wien2k_2001}


\section{\label{Results}Results}

\begin{figure}
\includegraphics[width=3.5in]{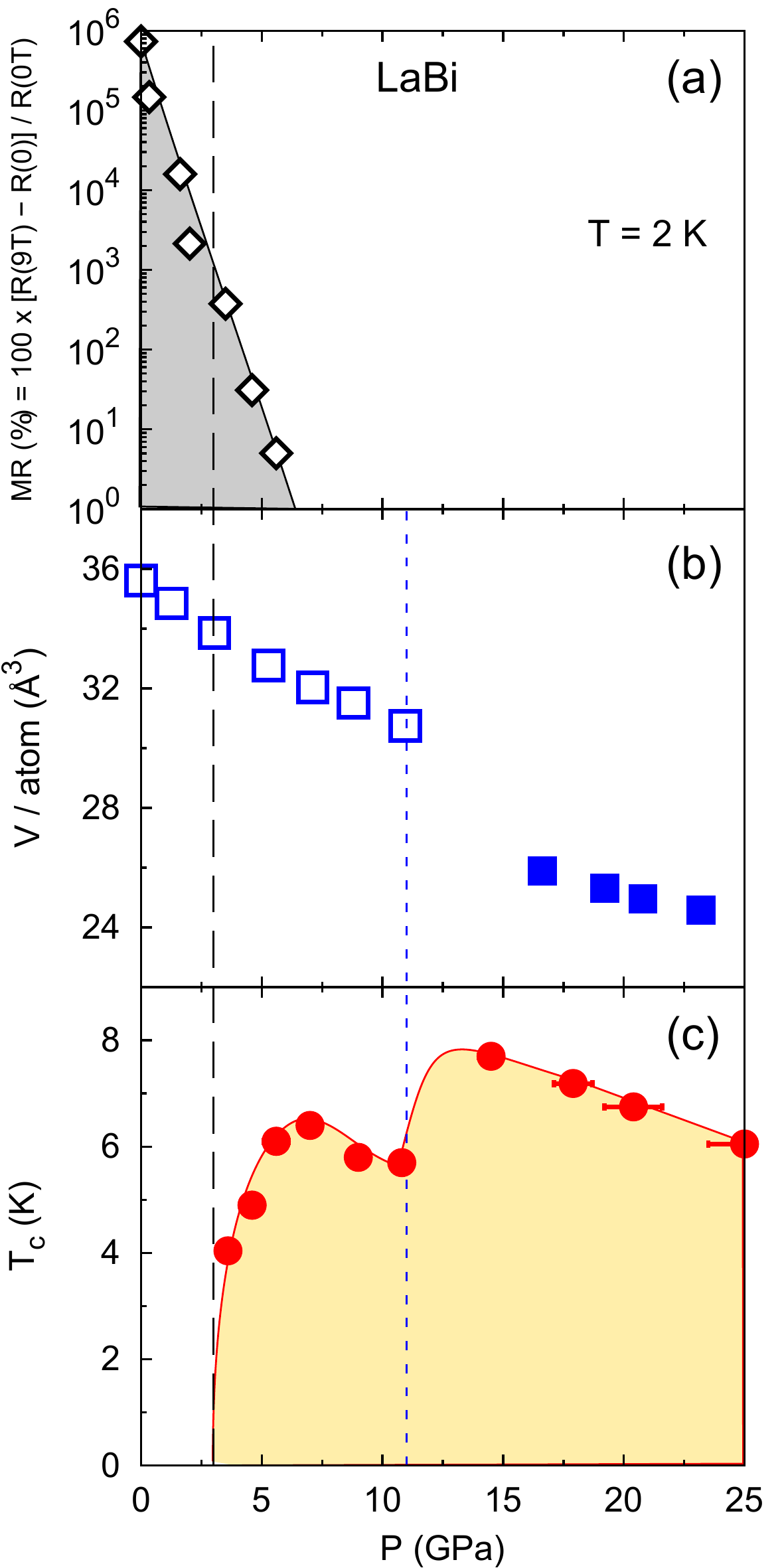}
\caption{\label{PD} 
(a) Magnetoresistance as a function of pressure in LaBi . 
The extreme magnetoresistance (XMR) is suppressed by $P\approx 5$ GPa.
(b) Unit cell volume per atom as a function of pressure.
Pressure reduces the cubic unit cell volume smoothly with no discontinuity across the region of XMR suppression. 
The discontinuous jump at $P\approx 11$ GPa is a structural transition from cubic to tetragonal marked by the vertical blue dotted line.
(c) Temperature-pressure phase diagram of superconductivity in LaBi. 
The onset of superconductivity is marked by the vertical black dashed line.
$T_c$ increases rapidly with increasing pressure until $P=6$ GPa, then decreases until $P=11$ GPa where it shows a sudden $40 \%$ increase concurrent with the structural transition.
For $P<15$ GPa, errorbars are no larger than the size of the data points.
}
\end{figure}

Fig. \ref{PD} summarizes our main findings and provides a guide for the rest of the article.
Fig. \ref{PD}(a) shows the suppression of magnetoreistance $\textrm{MR}=100\times \frac{R(9T)-R(0)}{R(0T)}$ by pressure.
Fig. \ref{PD}(b) shows smooth compression of the cubic unit cell with no structural anomaly as the extreme magnetoresistance (XMR) is suppressed by pressure.
Therefore, the suppression of XMR is due to smooth changes in the electronic structure of LaBi.
At $P \approx 11$ GPa a discontinuity occurs in the unit cell volume due to a structural transition.
Fig. \ref{PD}(c) shows that bulk superconductivity ($R=0$) starts at $P \approx 3.5$ GPa where XMR is substantially but not completely suppressed.
The discontinuity in the pressure dependence of $T_c$ at $P \approx 11$ GPa is due to the structural transition.
In the rest of the paper, we elaborate the effect of pressure on magnetoresistance, crystal structure, and superconductivity in LaBi.

\subsection{\label{extremeMR} The effect of pressure on XMR}

\begin{figure*}
\includegraphics[width=7in]{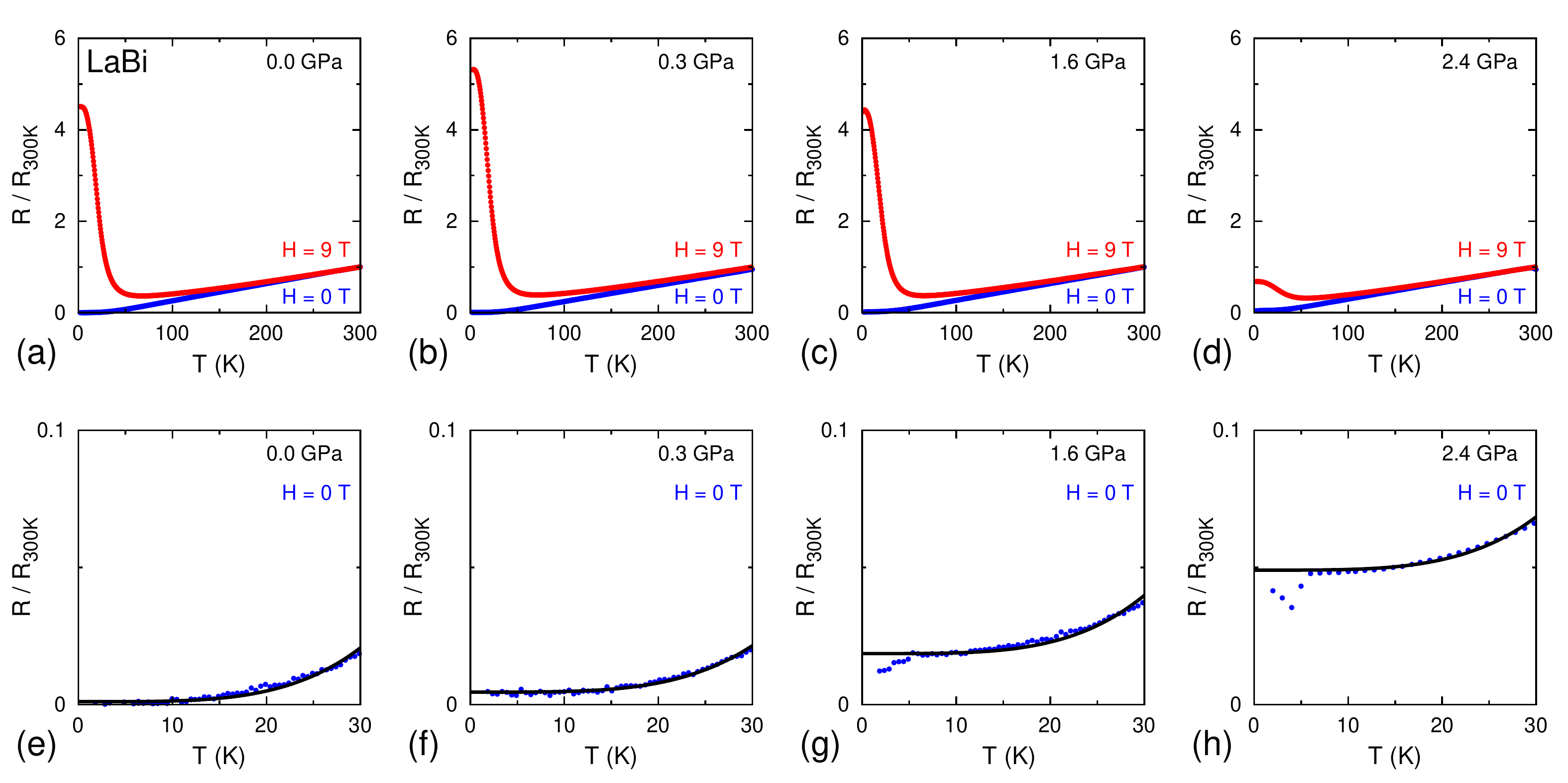}
\caption{\label{LP_RT_SC} 
Normalized resistance $R/R(300\textrm{K})$ at $H=0$ (blue) and $H=9$ T (red) as a function of temperature at $P=0$ (a), $P=0.3$ GPa (b), $P=1.6$ GPa (c), and $P=2.4$ GPa (d).
The normalized resistance at $H=9$ T in the plateau region is not systematically suppressed by increasing pressure and therefore does not explain the systematic suppression of XMR with pressure.
%
%
A zoom into the normalized resistance $R/R(300\textrm{K})$ at $H=0$ is shown for $T<30$ K at $P=0$ (e), $P=0.3$ GPa (f), $P=1.6$ GPa (g), and $P=2.4$ GPa (h).
Solid black lines are power law fits of the form $\rho=\rho_0 + AT^4$.
The zero-field resistance systematically increases with increasing pressure which explains the systematic suppression of XMR.
Broad and incomplete superconducting transitions appear at $P=1.6$ and 2.4 GPa, most likely due to pressure inhomogeneity. 
}
\end{figure*}

This section presents our data at lower pressures ($P<3$ GPa), from clamp cell experiments, to focus on the suppression of XMR with pressure.
Figs. \ref{LP_RT_SC}(a-d) compare the normalized resistance $R(T)/R(300 \textrm{K})$, at $H=0$ (blue) and $H=9$ T (red), at $P=0$, 0.3, 1.6, and 2.4 GPa.
The red curve in Fig. \ref{LP_RT_SC}(a) shows the typical profile of XMR with $\partial R / \partial T >0$ at $T>70$ K, $\partial R / \partial T <0$ at $20<T<70$ K, and $\partial R / \partial T \to 0$ at $T<20$ K.  
All topological semimetals with XMR show the same profile where $R(T)$ decreases initially with decreasing temperature, then increases, and finally saturates to a plateau. \cite{tafti_temperaturefield_2016}
The blue curve at $H=0$ shows metallic conduction where $R(T)$ decreases with decreasing temperature to a very small residual value.
Such small residual resistance $R(0)$ is essential to having an extremely large ratio $R(H)/R(0)$ \ie~XMR. 
Figs. \ref{LP_RT_SC}(a-d) show a moderate increase of $R(H=9\textrm{ T})$ in the \emph{plateau} region ($T<20$ K) from 0 to 0.34 GPa followed by a decrease at 1.6 GPa and a pronounced decrease at 2.4 GPa.
These changes do not account for the systematic suppression of XMR as a function of pressure shown in Fig. \ref{PD}(a).
To understand the systematic decrease of XMR we turn attention to the zero field resistance $R(H=0)$.
Figs. \ref{LP_RT_SC}(e-h) zoom into the normalized resistance at $H=0$ and $T<30$ K at $P=0$, 0.3, 1.6, and 2.4 GPa to reveal a systematic increase of the zero-field resistance by increasing pressure.
The black lines are fits to the expression $\rho=\rho_0+A T^4$ at each pressure.
The systematic increase of the zero-field resistance in Fig. \ref{LP_RT_SC}(e-h) explains the systematic decrease of XMR as a function of pressure in Fig. \ref{PD}(a).

\begin{figure*}
\includegraphics[width=7in]{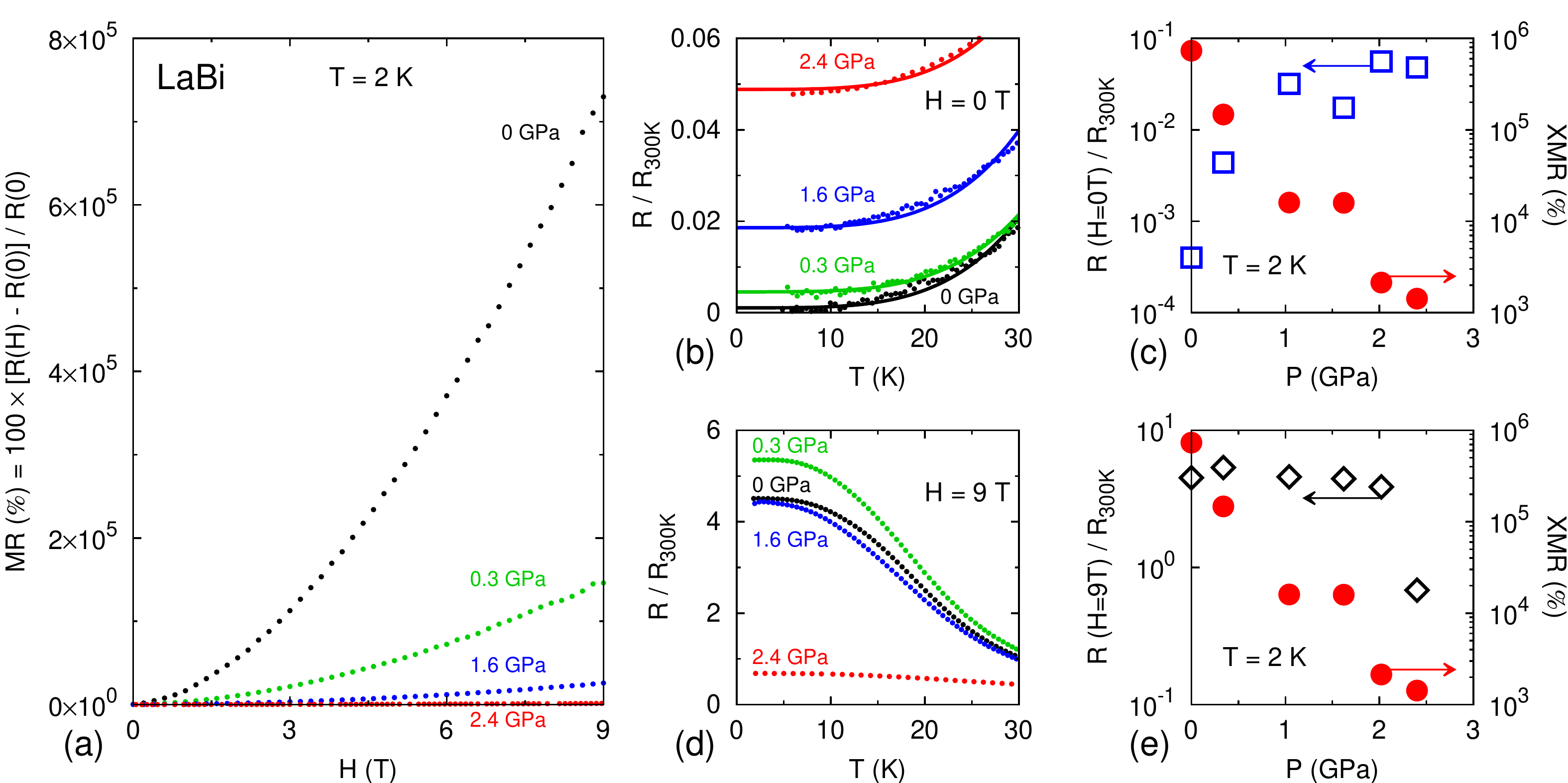}
\caption{\label{XMR} 
(a) Magnetoresistance MR $=100\times \frac{R(H)-R(0)}{R(0)}$ at $T=2$ K plotted as a function of field from $H=0$ to 9 T at four representative pressure values.
A systematic decrease of MR is observed with increasing pressure.
(b) Normalized resistance $R/R(300\textrm{ T})$ as a function of temperature at $H=0$ from $T=5$ to 30 K.
Solid lines are power law fits to the data at each pressure.
A systematic increase of the normalized resistance is observed with increasing pressure.
(c) Normalized resistance at $H=0$ and $T=2$ K are extracted from the fits in (b) and plotted as a function of pressure (empty blue squares corresponding to the left y-axis).
XMR values at $H=9$ T and $T=2$ K are extracted from (a) and plotted as a function of pressure (red circles corresponding to the right y-axis).
Both y-axes are logarithmic to show that the two quantities anti-correlate as they vary by orders of magnitude.
(d) Normalized resistance $R/R(300\textrm{ T})$ as a function of temperature from $T=2$ to 30 K at $H=9$ T.
(e) Normalized resistance at $H=9$ T and $T=2$ K are plotted as a function of pressure (empty black diamonds corresponding to the left y-axis).
XMR values at $H=9$ T and $T=2$ K are extracted from (a) and plotted as a function of pressure (red circles corresponding to the right y-axis).
There is no clear correlation between the two quantities.
}
\end{figure*}

Fig. \ref{XMR}(a) visualizes the suppression of XMR with pressure by plotting MR $= 100\times \frac{R(H)-R(0)}{R(0)}$ at $T=2$ K as a function of field at $P=0$, 0.3, 1.6, and 2.4 GPa.
Fig. \ref{XMR}(b) highlights the systematic increase of $R(0)$ at $H=0$ with increasing pressure. 
%
Fig. \ref{XMR}(c) shows a clear anti-correlation between increasing $R(0)$ and decreasing magnetoresistance. 
Both the left and the right y-axes are in logarithmic scale to compare the two quantities on equal footing.
In contrast, Figs. \ref{XMR}(d) and \ref{XMR}(e) show the absence of a clear correlation between $R(9\textrm{ T})$ and XMR.
%
%
Comparing Figs. \ref{XMR}(c) and \ref{XMR}(e) makes a compelling case that the zero-field resistance controls the magnitude of XMR in agreement with previous works that correlate the residual resistivity of various LaBi, LaSb, or WTe$_2$ samples with the magnitude of XMR. \cite{tafti_resistivity_2016, tafti_temperaturefield_2016, ali_correlation_2015}
It has been proposed that a combination of orbital mixing, Fermi surface sizes, and Fermi surface shapes is responsible for the extremely small $R(0)$ in LaBi and other topological semimetals. \cite{zeng_compensated_2016, tafti_temperaturefield_2016}
Pressure is an effective tool to tune all of these parameters and therefore tuning the residual resistance of LaBi systematically as shown in Fig. \ref{XMR}(b).

\subsection{\label{structure} The effect of pressure on the structure}

\begin{figure}
\includegraphics[width=3.5in]{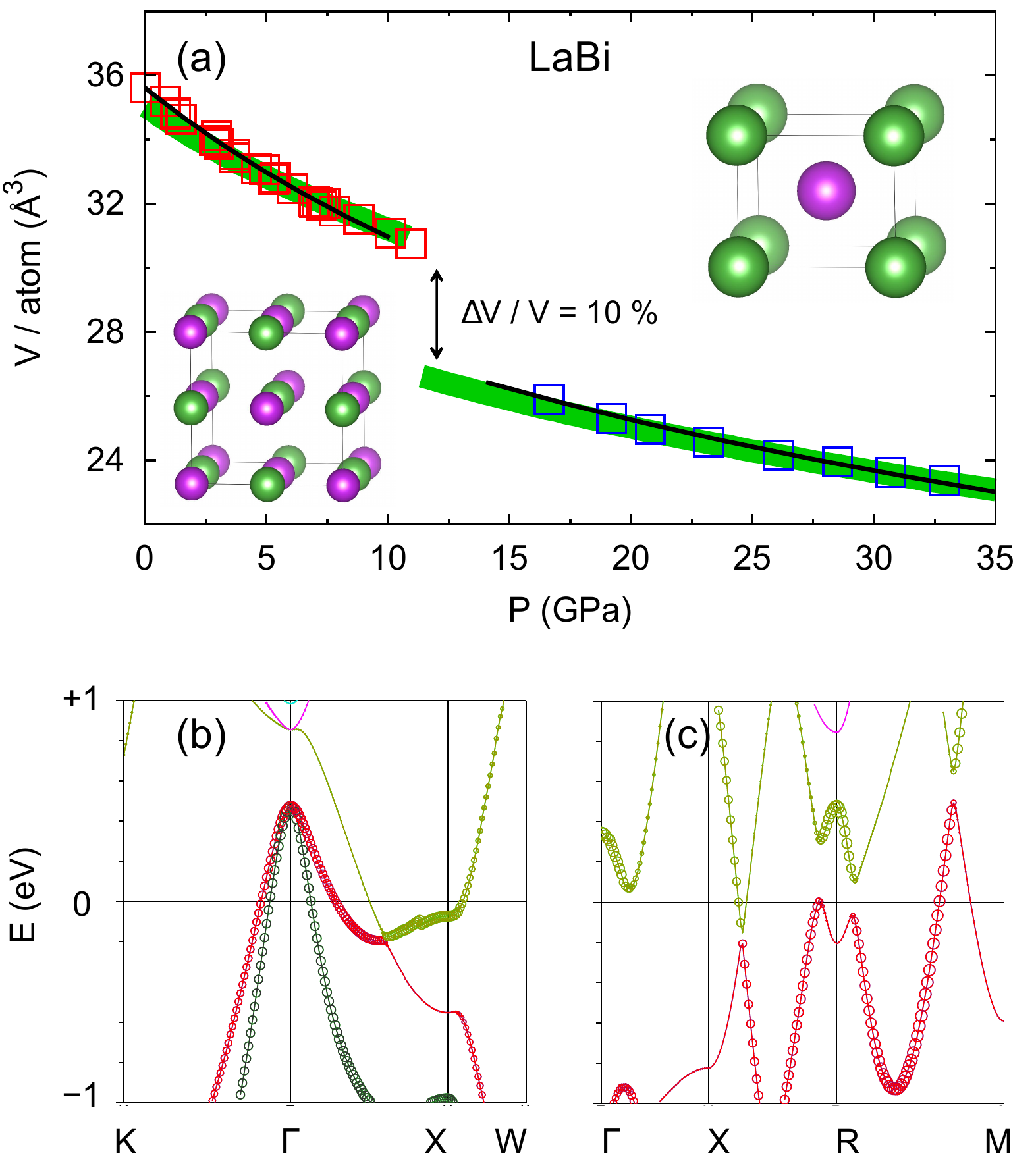}
\caption{\label{XRD_BS} 
(a) Unit cell volume per atom in LaBi as a function of pressure. 
The discontinuous drop at $P\approx 11$ GPa corresponds to a structural phase transition from face center cubic (FCC) to primitive tetragonal (PT) lattice as illustrated on the figure. 
Thick green lines are the results of theoretical calculations by Vaitheeswaran \etal. \cite{vaitheeswaran_electronic_2002}
Solid black lines are fits to the Birch-Murnaghan equation of state (Eq. \ref{Murnaghan}) from which we extract the bulk moduli for both structures as reported in table \ref{T1}.
Representative refinements are given in Appendix \ref{APP1}.
(b) Band structure of LaBi in the low pressure FCC structure with two central hole-pockets at $\Gamma$ and one electron-pocket at $X$.
(c) Band structure of LaBi in the high pressure PT structure with a small electron-pocket at $X$, a larger hole-pocket at $M$, and a gap with band inversion at $R$.
}
\end{figure}

Fig. \ref{XRD_BS}(a) shows that the unit cell volume of LaBi smoothly decreases with increasing pressure until $P\approx 11$ GPa.
There is no structural anomaly at lower pressures where extreme magnetoresistance is suppressed.
At 11 GPa there is a discontinuous $10 \%$ drop in the unit cell volume due to a structural transition from the face centered cubic lattice (FCC, space group $Fm\bar{3}m$) to a primitive tetragonal lattice (PT, space group $P4/mmm$).
The FCC to PT phase transition in LaBi has been theoretically predicted \cite{charifi_phase_2008, vaitheeswaran_electronic_2002, cui_first-principles_2009} but experimentally unresolved until now.  
Fig. \ref{XRD_BS}(a) shows that the onset of the structural transition at $P\approx 11$ GPa observed experimentally agrees with the theoretical predictions (thick green lines).
Solid black lines in Fig. \ref{XRD_BS}(a) are fits to the Birch-Murnaghan equation of state: \cite{murnaghan_finite_1937, birch_finite_1947}
\begin{equation}
\begin{split}
P(V) = \frac{3B}{2}\left[ \left( \frac{V_0}{V} \right)^{\frac{7}{3}} -  \left( \frac{V_0}{V} \right)^{\frac{5}{3}} \right] \times \\
\left[ 1 + \frac{3}{4} (B'-4)  \left[ \left( \frac{V_0}{V} \right)^{\frac{2}{3}} -  1 \right]   \right]
\label{Murnaghan}
\end{split}
\end{equation}
where $P_0$ and $V_0$ are the coordinates of the first data points in the FCC and the PT phases. The bulk modulus $B$ and its pressure derivative $B'=\partial B/ \partial P$ in the low pressure and the high pressure structures are extracted from the fits to Eq. \ref{Murnaghan} and summarized in table \ref{T1}.
In the low pressure FCC structure, our experimental value for the bulk modulus agrees with the theoretical calculations by Cui \etal \cite{cui_first-principles_2009} and Vaitheeswaran \etal \cite{vaitheeswaran_electronic_2002}.
In the high pressure PT structure, the two theory groups disagree.
Cui \etal~predict comparable bulk moduli between the low and the high pressure structures.
Vaitheeswaran \etal~predict a two-fold increase of the bulk modulus in the high pressure PT structure. 
Our data clearly agrees with the latter (see table \ref{T1}).
Representative powder x-ray diffraction data under pressure with Rietveld refinements are shown in Appendix \ref{APP1} for both FCC and PT phases.

\begin{table}[b]
\caption{\label{T1} 
The bulk modulus $B$ and its pressure derivative $B'=\partial B/ \partial P$ for LaBi extracted by fitting the data to the Birch-Murnaghan equation of state (Eq. \ref{Murnaghan}) as shown in Fig. \ref{XRD_BS}(a). The initial parameters $P_0$ and $V_0$ were fixed based on the experimental data in the low pressure face centered cubic (FCC) and the high pressure primitive tetragonal (PT) structures. 
}
\begin{ruledtabular}
\begin{tabular}{ccccc}
Bravais Lattice   & $B$ (GPa) & $B'$ & $P_0$ (GPa) & $V_0$ ($\AA$)\\
\colrule 
FCC  & $52 \pm 1 $ & $5.0 \pm 0.4$ & $0$ & $35.61$\\   
PT & $97 \pm 5$ & $5.8 \pm 0.9$ & $16.6$ & $25.90$

\end{tabular}
\end{ruledtabular}
\end{table}

The structural transition at 11 GPa changes the band structure of LaBi as shown in Figs. \ref{XRD_BS}(b) and \ref{XRD_BS}(c).
Fig. \ref{XRD_BS}(b) shows the band structure of LaBi in the low pressure FCC structure with two hole-pockets at the Brillouin zone center $\Gamma$ and one electron-pocket at $X$.
The small circles represent lanthanum $d$-states and the large circles represent bismuth $p$-states.
The mixing between $d$ and $p$ states on the electron pocket at $X$ has been attributed to the extremely small $R(0)$ and the large $R(H)$ in LaBi. \cite{tafti_temperaturefield_2016, zeng_topological_2015} 
The combination of orbital mixing, small ellipsoidal pockets, and electron-hole compensation as shown in Fig. \ref{XRD_BS}(b) is common to all topological semimetals and possibly the source of XMR. \cite{tafti_temperaturefield_2016} 
Fig. \ref{XRD_BS}(c) shows the electronic structure of LaBi in the high pressure PT phase with two notable changes compared to the low pressure FCC phase:
(1) The hole-pocket near $M$ is clearly larger than the electron-pocket near $X$ and therefore electron-hole compensation is weaker in the PT phase. 
%
The lack of electron-hole compensation in the PT phase explains the lack of magnetoresistance at high pressures.
(2)  There is a band inversion at the $R$ point with a gap due to the spin-orbit coupling.
Based on the Fu-Kane-Mele formula\cite{fu_topological_2007}, this gap corresponds to a strong topological insulator. 
However, the hole-pocket that crosses $E_F$ near $M$ prevents LaBi from being an insulator. 
The detailed evolution of the band structure in LaBi under pressure is given in Appendix \ref{APP2}.

\subsection{\label{superconductivity} The effect of pressure on superconductivity}

\begin{figure}
\includegraphics[width=3.5in]{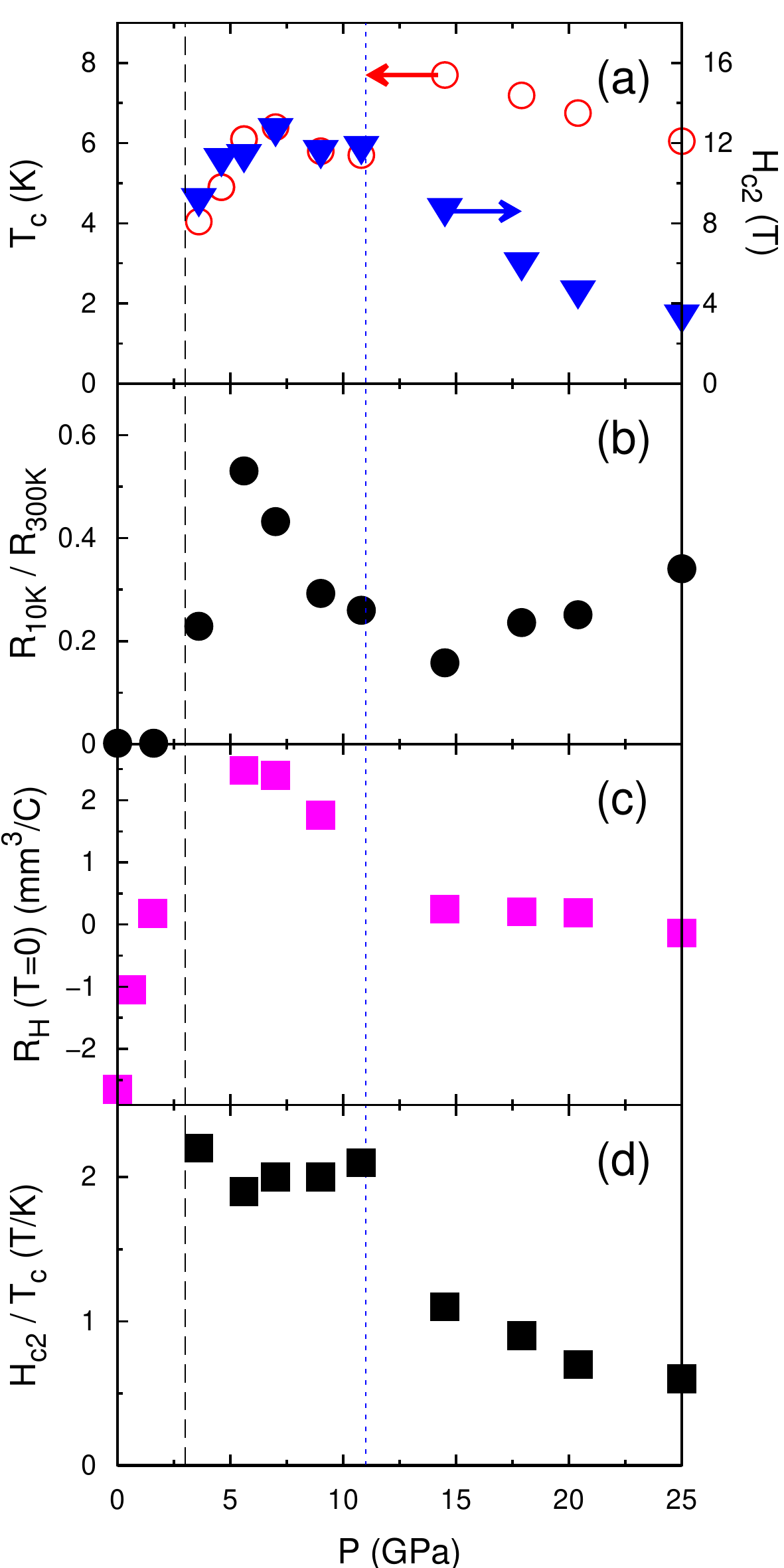}
\caption{\label{SC} 
(a) $T_c$ and $H_{c2}$ of LaBi as a function of pressure. 
Superconductivity onsets at $P\approx 3.5$ GPa, then $T_c$ shows $40 \%$ enhancement at the structural transition at $P\approx 11$ GPa.
(b) $R_{10{\textrm K}}/R_{300{\textrm K}}$ plotted as a function of pressure in LaBi.
The increase of $R_{10{\textrm K}}/R_{300{\textrm K}}$ is associated with the suppression of XMR.
At the structural transition ($P=11$ GPa), $R_{10{\textrm K}}/R_{300{\textrm K}}$ reverses direction from decreasing to increasing.
(c) The zero temperature limit of the Hall coefficient $R_H$ as a function of pressure.
There is a sign change from negative to positive at low pressures in the region of XMR suppression just before superconductivity appears.
$R_H$ falls to almost zero at the structural transition at $P\approx11$ GPa.
(d) The ratio $H_{c2}/T_c$ plotted as a function of pressure shows a sudden two-fold drop across the structural transition. 
The low pressure superconducting phase is not Pauli-limited but the high pressure phase is. 
}
\end{figure}

\begin{figure}
\includegraphics[width=3.5in]{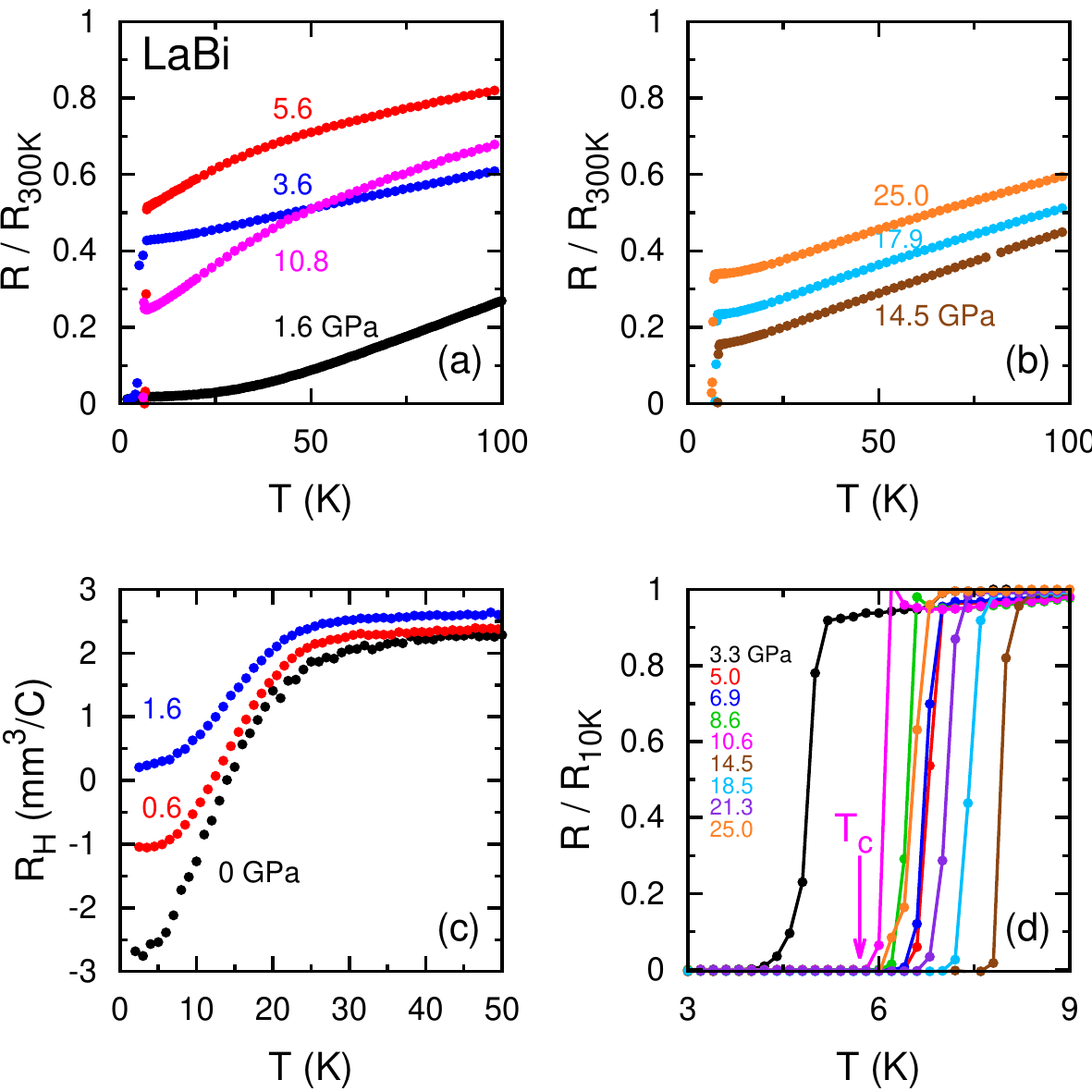}
\caption{\label{RES} 
(a) Normalized electrical resistance at $H=0$ T, from $T=2$ to 300 K, at several pressures below 11 GPa before the structral transition.
(b) Normalized electrical resistance at $H=0$ T, from $T=2$ to 300 K, at several pressures above 11 GPa after the structral transition.
(c) Hall coefficient as a function of temperature for several representative pressures. Note the sign change in $R_H (T=0)$ with increasing pressure.
(d) Normalized resistivity curves in the region of superconducting transition from which the phase diagrams in Figs. \ref{PD}
and \ref{SC} are constructed.
The arrow on the 10.6 GPa curve shows that we define $T_c$ using $R=0$ criterion.
}
\end{figure}

Fig. \ref{SC}(a) shows that the first complete superconducting transition ($R=0$) appears at $P \approx 3.5$ GPa in LaBi.
At this pressure, XMR is reduced by three orders of magnitude but not completely vanished as shown in Fig. \ref{PD}(a).
The onset of superconductivity is accompanied by two other observations, marked by the vertical black dashed line on Fig. \ref{SC}.
First, the normalized low temperature resistance ($R_{10{\textrm K}}/R_{300{\textrm K}}$) shows considerable increase at the onset of superconductivity (also see Fig. \ref{SC}(b)).
Second, the Hall coefficient ($R_H$) changes sign (Fig. \ref{SC}(c)).
The complete temperature profiles of resistivity and Hall data are presented in Fig. \ref{LP_RT_SC} and Fig. \ref{RES}.
A change of sign in $R_H$ concurrent with superconductivity was recently reported in another XMR material WTe$_2$. \cite{kang_superconductivity_2015}
In Appendix \ref{APP2} we use the experimental lattice parameters of LaBi to calculate the evolution of its band structure by increasing pressure.
Fig. \ref{BS_B1B2} in Appendix \ref{APP2} shows that the electron pocket size reduces with pressure in agreement with the change of sign in $R_H$ from negative to positive with increasing pressure as shown in Figs. \ref{SC}(c) and \ref{RES}(c).
Ref. \onlinecite{tafti_temperaturefield_2016} argues that the electron pocket plays a central role in XMR which is consistent with our observation of simultaneous suppression of XMR, sign change in $R_H$, and the appearance of superconductivity. 
The vertical blue dotted line on Fig. \ref{SC} marks the onset of structural transition at $P=11$ GPa as discussed in section \ref{structure}.
Due to the structural transition, $T_c$ shows a $40 \%$ increase (Fig. \ref{SC}(a)), $R_{10{\textrm K}}/R_{300{\textrm K}}$ reverses direction from decreasing to increasing (Fig. \ref{SC}(b)), and $R_H$ drops to almost zero (Fig. \ref{SC}(c)). 
The complete $R(T)$ profiles are presented in Figs. \ref{RES}(a) and \ref{RES}(b).
Such drastic changes in transport properties follow the drastic change of band structure as a result of the structural transition shown in Fig. \ref{XRD_BS}.

\begin{figure}
\includegraphics[width=3.5in]{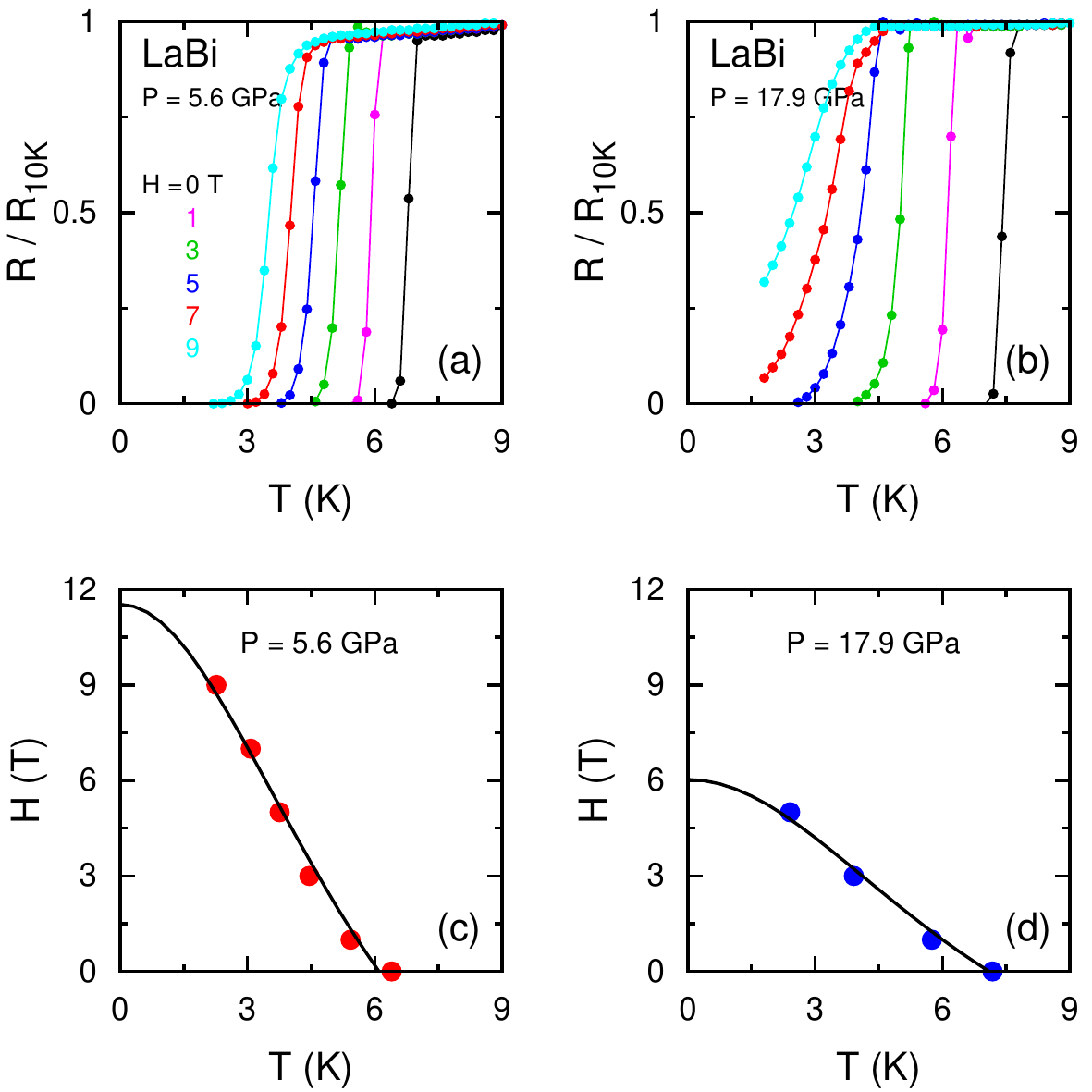}
\caption{\label{Hc2} 
(a) Normalized electrical resistance at $P=5.6$ GPa plotted as a function of temperature in several magnetic fields as indicated on the figure.
(b) Normalized electrical resistance at $P=17.9$ GPa plotted as a function of temperature in several magnetic fields.
(c) The $H$-$T$ data for the superconducting transition from (a) are fitted to Eq. \ref{WHH} with the resulting $H_{c2}=11.5$ T.
(d) The $H$-$T$ data for the superconducting transition from (b) are fitted to Eq. \ref{WHH} with the resulting $H_{c2}=6.1$ T.
}
\end{figure}

Fig. \ref{SC}(a) shows both $T_c$ (left y-axis) and $H_{c2}$ (right y-axis) at each pressure.
The high values of $H_{c2}$ and the onset of superconductivity at 3.5 GPa rule out Bismuth filamentary superconductivity which onsets at  8 GPa with $H_{c2} < 0.5$ T . \cite{ilina_1972, baring_local_2011}
Fig. \ref{Hc2} shows how we derive $H_{c2}$ of LaBi using the extended Ginzburg-Landau formalism \cite{zhu_upper_2008, fang_fabrication_2005}
\begin{equation}
H_{c2}(T)=H_{c2}(0)\frac{1-\left(T/T_c\right)^2}{1+\left(T/T_c\right)^2}
\label{WHH}
\end{equation}
where $H_{c2}(0)$ is the upper critical field at $T=0$.
The values of $H_{c2}=11.5$ T at $P=5.6$ GPa and $H_{c2}=6.1$ T at $P=17.9$ GPa are order of magnitude larger than the reported values of $H_{c2}$ in Bismuth. \cite{zhu_upper_2008, fang_fabrication_2005}
More details on Bi superconductivity are given in Appendix \ref{APP3}. 
Fig. \ref{SC}(a) shows that before the structural transition at $P=11$ GPa, $H_{c2}$ values are almost double the value of $T_c$ at each pressure. 
Since the Pauli limit of superconductivity is given by $H_{c2}=1.85 T_c$, the superconducting phase below 11 GPa seems not to be Pauli-limited.
Interestingly, for $P>11$ GPa \ie~after the structural transition, the ratio of $H_{c2}/T_c$ suddenly drops to near unity.
Fig. \ref{SC}(d) traces the $H_{c2}/T_c$ ratio as a function of pressure revealing the sudden transition from not-Pauli-limited superconductivity to Paul-limited superconductivity across the structural transition.
Spin triplet pairing is possible in the low pressure not-Pauli-limited phase but not in the high pressure Paul-limited phase. 
Recent studies show a change of $T_c$ with structural transition in ZrTe$_5$. \cite{zhou_pressure-induced_2016}
It would be interesting to look for the same effects in ZrTe$_5$ and other XMR materials which superconduct under pressure. 

\section{\label{Summary}Summary}

In summary, we study the effect of pressure on extreme magnetoresistance, crystal structure, and superconducting properties of LaBi.
Pressure suppresses XMR and gives rise to superconductivity in LaBi (Fig. \ref{PD}).
The suppression of XMR anti-correlates with the increase of the residual resistance $R(0)$ as shown in Fig. \ref{XMR}(c).
It does not correlate with the in-field resistance $R({9\textrm T})$ as shown in Fig. \ref{XMR}(e).
The suppression of XMR is accompanied by a sign change in the Hall coefficient $R_H$ from negative to positive as shown in Figs. \ref{SC}(c) and \ref{RES}(c).
This is consistent with the recent argument that the electron Fermi surface with orbital mixing is responsible for the extremely small $R(0)$ and therefore XMR. \cite{tafti_temperaturefield_2016}
Our DFT calculations in Fig. \ref{BS_B1B2} in Appendix \ref{APP2} confirm that the $R_H$ sign change is due to the shrinking of the electron pocket with increasing pressure. 
The pressure induced structural transition in LaBi from FCC to PT shown in Fig. \ref{XRD_BS} is a new finding.
The change in the crystal structure changes the band structure and creates a region of band inversion in LaBi (Fig. \ref{XRD_BS}).
The changes in the band structure of LaBi due to this structural transition give rise to a reversal in $R_{10{\textrm K}}/R_{300{\textrm K}}$ from decreasing to increasing and a drop in $R_H$ as shown in Figs. \ref{SC}(b) and \ref{SC}(c).
At the structural transition, there is a discontinuity in $T_c$ showing an abrupt $40 \%$ increase (Fig. \ref{SC}(a)).
Simultaneously, the ratio $H_{c2}/T_c$ shows a step-like change from $H_{c2}/T_c \approx 2$ \ie~not-Pauli-limited superconductivity to $H_{c2}/T_c \approx 1$ \ie~Pauli-limited superconductivity as shown in Fig. \ref{SC}(d).
%


\section*{ACKNOWLEDGMENTS}

This work was performed under LDRD (Tracking Code No. 14-ERD-041) and under the auspices of the U.S. Department of Energy (DOE) by Lawrence Livermore National Laboratory (LLNL) under Contract No. DE-AC52- 07NA27344. Portions of this work were performed at HPCAT (Sector 16), Advanced Photon Source (APS), Argonne National Laboratory. HPCAT operations are supported by the DOE-NNSA under Award No. DE-NA0001974 and the DOE-BES under Award No. DE-FG02-99ER45775 with partial instrumentation funding by the NSF. The Advanced Photon Source is a U.S. DOE Office of Science User Facility operated for the DOE Office of Science by Argonne National Laboratory under Contract No. DE-AC02-06CH11357. Beam time was provided by the Carnegie DOE-Alliance Center (CDAC). Y.K.V. acknowledges support from DOE-NNSA Grant No. DE-NA0002014.
The research at Princeton was supported by the Gordon and Betty Moore Foundation under the EPiQS program, grant GBMF 4412 and the ARO MURI on topological insulators, grant W911NF-12-1-0461.  


\appendix

\section{\label{APP1}Rietveld refinement of high pressure XRD data}

\begin{figure}
\includegraphics[width=3.5in]{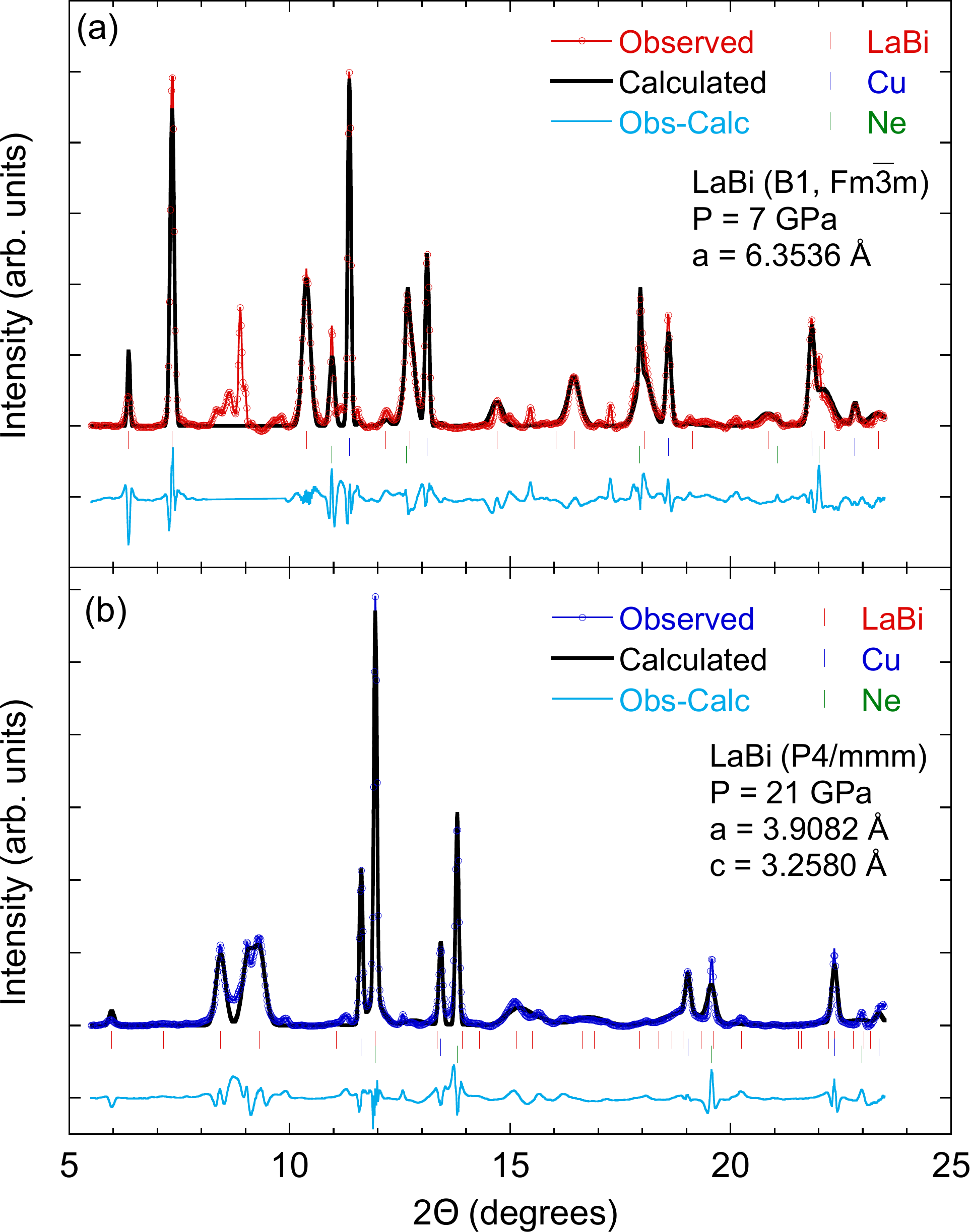}
\caption{\label{Refinements} 
Representative refinement of the x-ray diffraction patterns collected at (a) $P=7$~GPa and (b) $P=21$~GPa.
Empty circles show the XRD data plotted as intensity versus $2\Theta$.
Black lines are the best fit to the data.
Blue lines show the difference between the data and the fits.
Cu (pressure gauge) and Ne (pressure transmitting medium) peaks are indexed individually. 
}
\end{figure}

Fig. \ref{Refinements} includes two representative structural refinements of the x-ray diffraction data at $P=7$~GPa and $P=21$~GPa.
The low pressure structure is rock-salt (B1) and the high pressure structure has a primitive tetragonal unit cell as illustrated on Fig. \ref{XRD_BS}. 
In Fig. \ref{Refinements}(a), the peaks between 8 and 9 degrees have been excluded from the refinement, and they are likely to come from small inclusions of elemental Bi. 
At 7 GPa, Bi is in a complex host-guest structure, which is difficult to refine with so few evident peaks. 
For $P > 8$ GPa, elemental Bi is BCC, and we do include this phase in the refinement; the most prominent Bi peak occurs near 9 degrees in Fig 7b.

\section{\label{APP2}Evolution of LaBi band structure with pressure}

\begin{figure}
\includegraphics[width=3.5in]{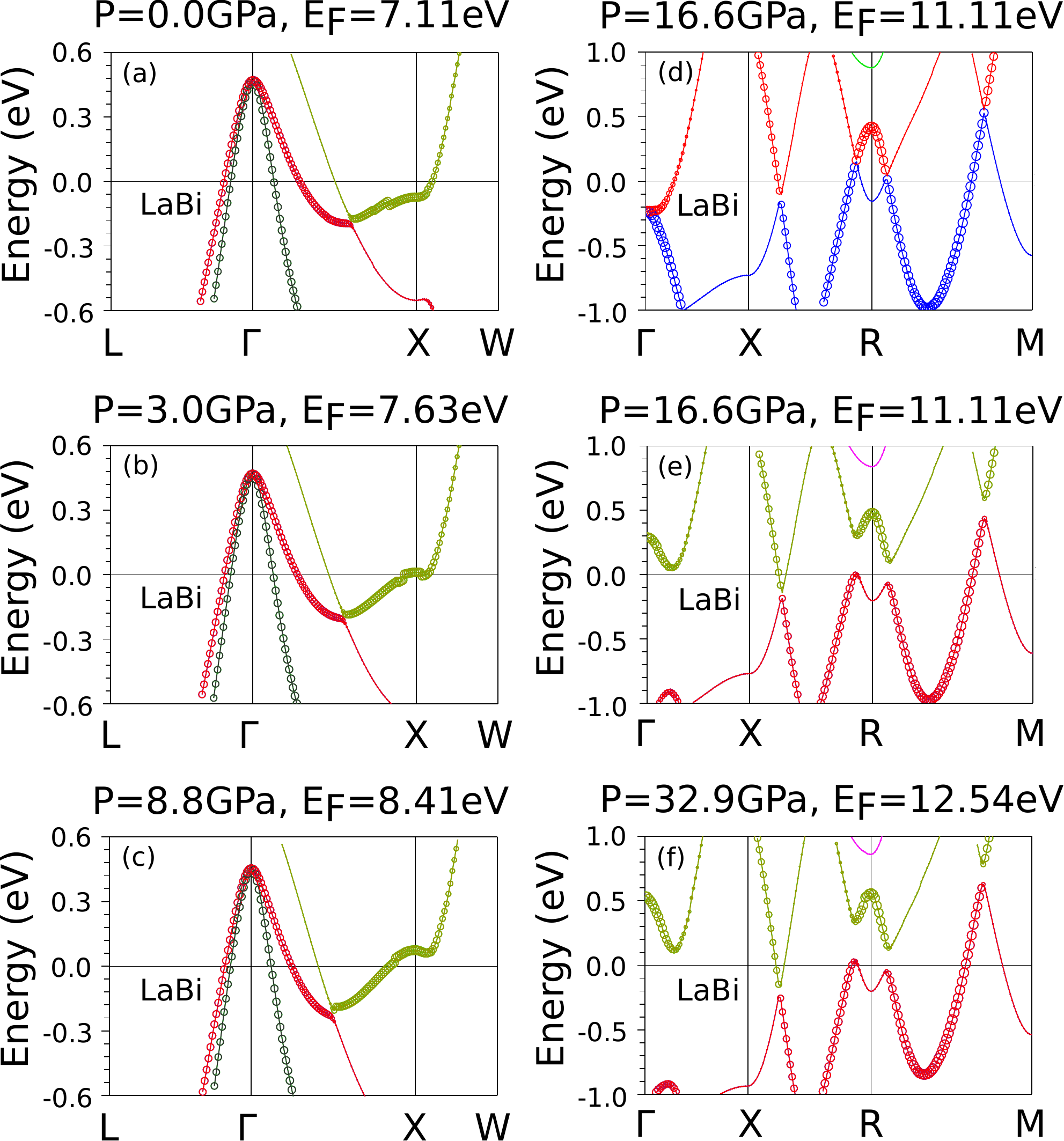}
\caption{\label{BS_B1B2} 
(a) Band structure of LaBi at $p=0$ in the FCC structure. 
The large circles represent the $p$-orbitals of Bi and the small circles represent the $d$-orbitals of La.
The y-axis is energy relative to $E_F$ with the $E_F$ given on top of each panel.
(b) Band structure of LaBi at $p=3.0$ GPa. 
Notice that the electron pocket shrunks in size and its shape changes from cylindrical to round. 
(c) Band structure of LaBi at $p=8.8$ GPa.
The electron pocket continues to shrink and become more spherical.
Notice that the Fermi energy $E_F$ increases with increasing pressure and makes the material less compensated.
(d) Band structure of LaBi at $p=16.6$ GPa in the PT structure after the structural transition.
This calculation is without spin-orbit coupling to show the mixing between the bands at $R$.
(e) After including SOC, the bands hybridize and a gap opens at $R$ with a clear band inversion.
(f) Band structure of LaBi at $p=32.9$ GPa in the PT structure.
Pressure does not change the band structure that much in this phase.
}
\end{figure}

Figs. \ref{BS_B1B2}(a-c) present the band structure of LaBi in the FCC structure at $P=0$, 3.0, and 8.8 GPa before the structural transition.
Larger circles represent Bi $p$-states and smaller circles represent La $d$-states.
The calculation is based on our experimental values for the lattice parameters of LaBi (see Fig. \ref{XRD_BS}).
With increasing pressure, the Fermi energy $E_F$ increases and the size of the electron pocket at $X$ shrinks.
Pressure also changes the shape of this pocket from cigar-shape to round.
Figs. \ref{BS_B1B2}(d-f) present the band structure in the PT structure after the structural transition at $P=16.6$ and 32.9 GPa.
Fig. \ref{BS_B1B2}(d) shows the results of DFT calculations in the PT structure before including spin-orbit coupling.
As a result of SOC, the two bands that cross at $R$ will hybridize to form a band inverted gap as shown in Fig. \ref{BS_B1B2}(e).
Increasing pressure in the PT phase does not change the band structure visibly as shown in Fig. \ref{BS_B1B2}(f) which is due to the stiffer structure in the PT phase (table \ref{T1}).
The band structure plotted in Fig. \ref{BS_B1B2}(a) gives rise to the extreme magnetoresistance and a negative $R_H$ in LaBi, in (b) XMR is reduced, $R_H$ has changed sign to positive, and the material is on the verge of becoming a superconductor, in (c) XMR is completely gone and the material is superconducting in the FCC structure with $H_{c2}/T_c\approx 2$, in (e) the material has gone through the structural phase transition, it continues to superconduct in the PT structure but with $H_{c2}/T_c\approx 1$, in (f) LaBi is still superconducting in the PT phase with $R_H$ becoming nearly zero.

\section{\label{APP3}Superconductivity in LaBi and Bi}

\begin{figure}
\includegraphics[width=3.5in]{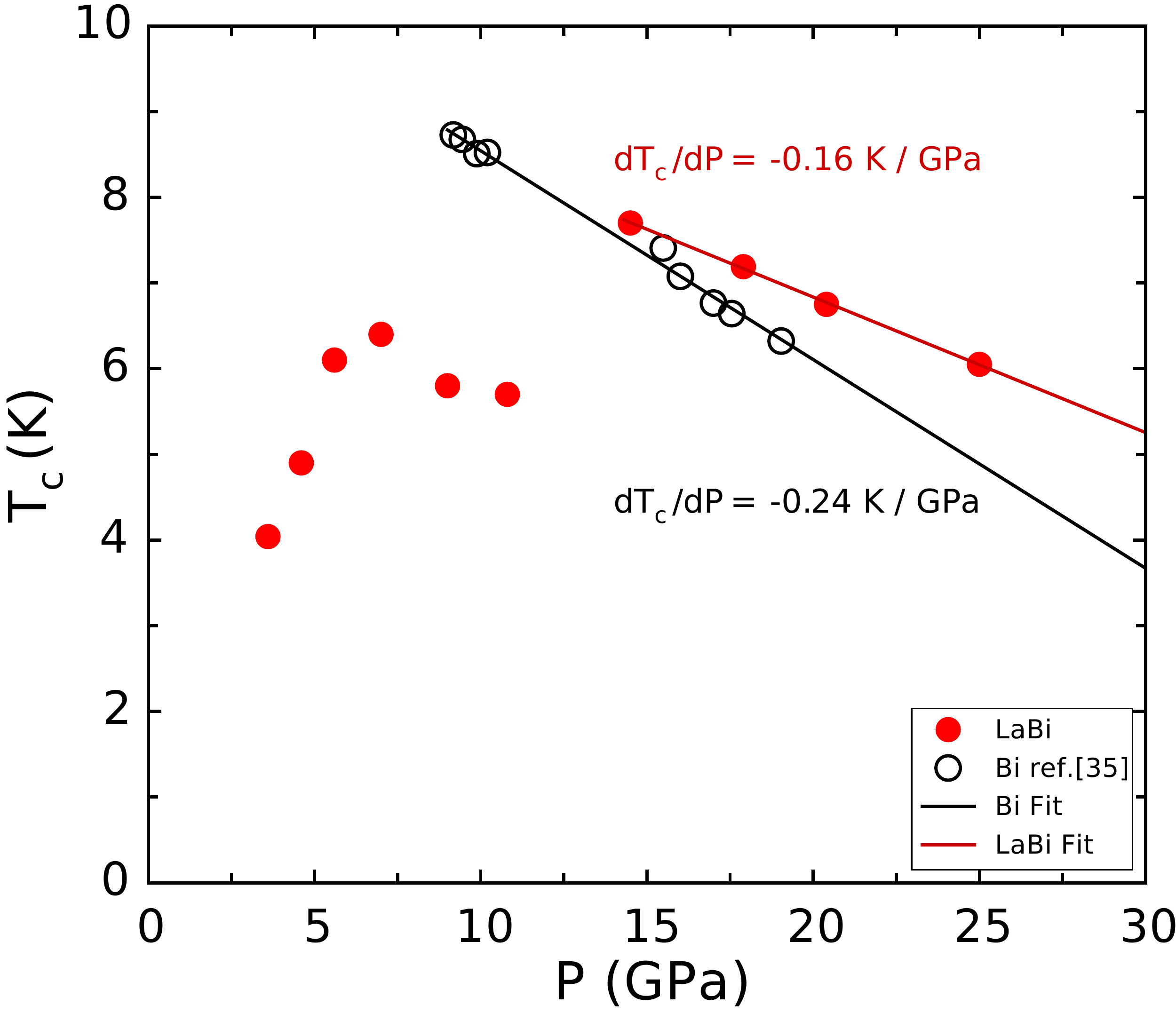}
\caption{\label{Bi_LaBi} 
$T_c$ plotted as a function of pressure in LaBi (solid red circles) and Bi (open black circles).
Data points for Bi come from Ref. \onlinecite{ilina_1972}.
Superconductivity in Bi starts from $P=8$ GPa with higher $T_c$ than LaBi at the same pressure.
We do not see short circuiting due to Bi superconductivity. 
Also note that the pressure slope of $T_c$ in Bi is 1.5 times larger than LaBi.
}
\end{figure}

In the main text we mentioned two reasons to believe that the superconducting transitions we observe do not come from elemental Bismuth: (a) Superconductivity of Bi starts at lower pressures with higher $T_c$, and (b) it has a much lower $H_{c2}$. \cite{ilina_1972, baring_local_2011}
Fig. \ref{Bi_LaBi} shows the pressure dependence of $T_c$ in Bi from Ref. \onlinecite{ilina_1972} in open black symbols and compares it to our data in LaBi as full red symbols.
Superconductivity in Bi onsets at $P=8$ GPa with $T_c=8.9$ K.
At the same pressure, $T_c=5.8$ K in LaBi.
If the superconducting signal was due to elemental Bi, it should have short circuited the resistivity of LaBi and mask the transition at $5.8$ K.
Further, the pressure dependence of $T_c$ in LaBi at high pressures has a slope $dT_c/dP= - 0.16$ K/GPa compared to Bi with $dT_c/dP= - 0.24$ which is 1.5 times larger. 
Lastly, the $H_{c2}$ values we extract for LaBi as shown in Fig. \ref{Hc2} are a few teslas compared to $H_{c2},0.5$ T in elemental Bismuth. \cite{ilina_1972, baring_local_2011}
Therefore, the superconduting transitions observed here cannot come from Bi impurity phases.
LaBi is grown out of Indium flux which also superconducts with $T_c=3.4$ K at zero pressure.
We clearly do not see any superconducting transitions at zero pressure and the first complete transition appears at 3.5 GPa which cannot be due to indium because the $T_c$ of indium decreases with pressure.


\bibliography{LaBi_07july2016}

\begin{thebibliography}{38}%
\makeatletter
\providecommand \@ifxundefined [1]{%
 \@ifx{#1\undefined}
}%
\providecommand \@ifnum [1]{%
 \ifnum #1\expandafter \@firstoftwo
 \else \expandafter \@secondoftwo
 \fi
}%
\providecommand \@ifx [1]{%
 \ifx #1\expandafter \@firstoftwo
 \else \expandafter \@secondoftwo
 \fi
}%
\providecommand \natexlab [1]{#1}%
\providecommand \enquote  [1]{``#1''}%
\providecommand \bibnamefont  [1]{#1}%
\providecommand \bibfnamefont [1]{#1}%
\providecommand \citenamefont [1]{#1}%
\providecommand \href@noop [0]{\@secondoftwo}%
\providecommand \href [0]{\begingroup \@sanitize@url \@href}%
\providecommand \@href[1]{\@@startlink{#1}\@@href}%
\providecommand \@@href[1]{\endgroup#1\@@endlink}%
\providecommand \@sanitize@url [0]{\catcode `\\12\catcode `\$12\catcode
  `\&12\catcode `\#12\catcode `\^12\catcode `\_12\catcode `\%12\relax}%
\providecommand \@@startlink[1]{}%
\providecommand \@@endlink[0]{}%
\providecommand \url  [0]{\begingroup\@sanitize@url \@url }%
\providecommand \@url [1]{\endgroup\@href {#1}{\urlprefix }}%
\providecommand \urlprefix  [0]{URL }%
\providecommand \Eprint [0]{\href }%
\providecommand \doibase [0]{http://dx.doi.org/}%
\providecommand \selectlanguage [0]{\@gobble}%
\providecommand \bibinfo  [0]{\@secondoftwo}%
\providecommand \bibfield  [0]{\@secondoftwo}%
\providecommand \translation [1]{[#1]}%
\providecommand \BibitemOpen [0]{}%
\providecommand \bibitemStop [0]{}%
\providecommand \bibitemNoStop [0]{.\EOS\space}%
\providecommand \EOS [0]{\spacefactor3000\relax}%
\providecommand \BibitemShut  [1]{\csname bibitem#1\endcsname}%
\let\auto@bib@innerbib\@empty
\bibitem [{\citenamefont {Liang}\ \emph {et~al.}(2015)\citenamefont {Liang},
  \citenamefont {Gibson}, \citenamefont {Ali}, \citenamefont {Liu},
  \citenamefont {Cava},\ and\ \citenamefont {Ong}}]{liang_ultrahigh_2015}%
  \BibitemOpen
  \bibfield  {author} {\bibinfo {author} {\bibfnamefont {T.}~\bibnamefont
  {Liang}}, \bibinfo {author} {\bibfnamefont {Q.}~\bibnamefont {Gibson}},
  \bibinfo {author} {\bibfnamefont {M.~N.}\ \bibnamefont {Ali}}, \bibinfo
  {author} {\bibfnamefont {M.}~\bibnamefont {Liu}}, \bibinfo {author}
  {\bibfnamefont {R.~J.}\ \bibnamefont {Cava}}, \ and\ \bibinfo {author}
  {\bibfnamefont {N.~P.}\ \bibnamefont {Ong}},\ }\href {\doibase
  10.1038/nmat4143} {\bibfield  {journal} {\bibinfo  {journal} {Nature
  Materials}\ }\textbf {\bibinfo {volume} {14}},\ \bibinfo {pages} {280}
  (\bibinfo {year} {2015})}\BibitemShut {NoStop}%
\bibitem [{\citenamefont {Xiong}\ \emph {et~al.}(2015)\citenamefont {Xiong},
  \citenamefont {Kushwaha}, \citenamefont {Liang}, \citenamefont {Krizan},
  \citenamefont {Hirschberger}, \citenamefont {Wang}, \citenamefont {Cava},\
  and\ \citenamefont {Ong}}]{xiong_evidence_2015}%
  \BibitemOpen
  \bibfield  {author} {\bibinfo {author} {\bibfnamefont {J.}~\bibnamefont
  {Xiong}}, \bibinfo {author} {\bibfnamefont {S.~K.}\ \bibnamefont {Kushwaha}},
  \bibinfo {author} {\bibfnamefont {T.}~\bibnamefont {Liang}}, \bibinfo
  {author} {\bibfnamefont {J.~W.}\ \bibnamefont {Krizan}}, \bibinfo {author}
  {\bibfnamefont {M.}~\bibnamefont {Hirschberger}}, \bibinfo {author}
  {\bibfnamefont {W.}~\bibnamefont {Wang}}, \bibinfo {author} {\bibfnamefont
  {R.~J.}\ \bibnamefont {Cava}}, \ and\ \bibinfo {author} {\bibfnamefont
  {N.~P.}\ \bibnamefont {Ong}},\ }\href {\doibase 10.1126/science.aac6089}
  {\bibfield  {journal} {\bibinfo  {journal} {Science}\ }\textbf {\bibinfo
  {volume} {350}},\ \bibinfo {pages} {413} (\bibinfo {year}
  {2015})}\BibitemShut {NoStop}%
\bibitem [{\citenamefont {Shekhar}\ \emph {et~al.}(2015)\citenamefont
  {Shekhar}, \citenamefont {Nayak}, \citenamefont {Sun}, \citenamefont
  {Schmidt}, \citenamefont {Nicklas}, \citenamefont {Leermakers}, \citenamefont
  {Zeitler}, \citenamefont {Skourski}, \citenamefont {Wosnitza}, \citenamefont
  {Liu}, \citenamefont {Chen}, \citenamefont {Schnelle}, \citenamefont
  {Borrmann}, \citenamefont {Grin}, \citenamefont {Felser},\ and\ \citenamefont
  {Yan}}]{shekhar_extremely_2015}%
  \BibitemOpen
  \bibfield  {author} {\bibinfo {author} {\bibfnamefont {C.}~\bibnamefont
  {Shekhar}}, \bibinfo {author} {\bibfnamefont {A.~K.}\ \bibnamefont {Nayak}},
  \bibinfo {author} {\bibfnamefont {Y.}~\bibnamefont {Sun}}, \bibinfo {author}
  {\bibfnamefont {M.}~\bibnamefont {Schmidt}}, \bibinfo {author} {\bibfnamefont
  {M.}~\bibnamefont {Nicklas}}, \bibinfo {author} {\bibfnamefont
  {I.}~\bibnamefont {Leermakers}}, \bibinfo {author} {\bibfnamefont
  {U.}~\bibnamefont {Zeitler}}, \bibinfo {author} {\bibfnamefont
  {Y.}~\bibnamefont {Skourski}}, \bibinfo {author} {\bibfnamefont
  {J.}~\bibnamefont {Wosnitza}}, \bibinfo {author} {\bibfnamefont
  {Z.}~\bibnamefont {Liu}}, \bibinfo {author} {\bibfnamefont {Y.}~\bibnamefont
  {Chen}}, \bibinfo {author} {\bibfnamefont {W.}~\bibnamefont {Schnelle}},
  \bibinfo {author} {\bibfnamefont {H.}~\bibnamefont {Borrmann}}, \bibinfo
  {author} {\bibfnamefont {Y.}~\bibnamefont {Grin}}, \bibinfo {author}
  {\bibfnamefont {C.}~\bibnamefont {Felser}}, \ and\ \bibinfo {author}
  {\bibfnamefont {B.}~\bibnamefont {Yan}},\ }\href {\doibase 10.1038/nphys3372}
  {\bibfield  {journal} {\bibinfo  {journal} {Nature Physics}\ }\textbf
  {\bibinfo {volume} {11}},\ \bibinfo {pages} {645} (\bibinfo {year}
  {2015})}\BibitemShut {NoStop}%
\bibitem [{\citenamefont {Ghimire}\ \emph {et~al.}(2015)\citenamefont
  {Ghimire}, \citenamefont {Luo}, \citenamefont {Neupane}, \citenamefont
  {Williams}, \citenamefont {Bauer},\ and\ \citenamefont
  {Ronning}}]{ghimire_magnetotransport_2015}%
  \BibitemOpen
  \bibfield  {author} {\bibinfo {author} {\bibfnamefont {N.~J.}\ \bibnamefont
  {Ghimire}}, \bibinfo {author} {\bibfnamefont {Y.}~\bibnamefont {Luo}},
  \bibinfo {author} {\bibfnamefont {M.}~\bibnamefont {Neupane}}, \bibinfo
  {author} {\bibfnamefont {D.~J.}\ \bibnamefont {Williams}}, \bibinfo {author}
  {\bibfnamefont {E.~D.}\ \bibnamefont {Bauer}}, \ and\ \bibinfo {author}
  {\bibfnamefont {F.}~\bibnamefont {Ronning}},\ }\href {\doibase
  10.1088/0953-8984/27/15/152201} {\bibfield  {journal} {\bibinfo  {journal}
  {Journal of Physics: Condensed Matter}\ }\textbf {\bibinfo {volume} {27}},\
  \bibinfo {pages} {152201} (\bibinfo {year} {2015})}\BibitemShut {NoStop}%
\bibitem [{\citenamefont {Huang}\ \emph {et~al.}(2015)\citenamefont {Huang},
  \citenamefont {Zhao}, \citenamefont {Long}, \citenamefont {Wang},
  \citenamefont {Chen}, \citenamefont {Yang}, \citenamefont {Liang},
  \citenamefont {Xue}, \citenamefont {Weng}, \citenamefont {Fang},
  \citenamefont {Dai},\ and\ \citenamefont {Chen}}]{huang_observation_2015}%
  \BibitemOpen
  \bibfield  {author} {\bibinfo {author} {\bibfnamefont {X.}~\bibnamefont
  {Huang}}, \bibinfo {author} {\bibfnamefont {L.}~\bibnamefont {Zhao}},
  \bibinfo {author} {\bibfnamefont {Y.}~\bibnamefont {Long}}, \bibinfo {author}
  {\bibfnamefont {P.}~\bibnamefont {Wang}}, \bibinfo {author} {\bibfnamefont
  {D.}~\bibnamefont {Chen}}, \bibinfo {author} {\bibfnamefont {Z.}~\bibnamefont
  {Yang}}, \bibinfo {author} {\bibfnamefont {H.}~\bibnamefont {Liang}},
  \bibinfo {author} {\bibfnamefont {M.}~\bibnamefont {Xue}}, \bibinfo {author}
  {\bibfnamefont {H.}~\bibnamefont {Weng}}, \bibinfo {author} {\bibfnamefont
  {Z.}~\bibnamefont {Fang}}, \bibinfo {author} {\bibfnamefont {X.}~\bibnamefont
  {Dai}}, \ and\ \bibinfo {author} {\bibfnamefont {G.}~\bibnamefont {Chen}},\
  }\href {\doibase 10.1103/PhysRevX.5.031023} {\bibfield  {journal} {\bibinfo
  {journal} {Physical Review X}\ }\textbf {\bibinfo {volume} {5}},\ \bibinfo
  {pages} {031023} (\bibinfo {year} {2015})}\BibitemShut {NoStop}%
\bibitem [{\citenamefont {Wang}\ \emph {et~al.}(2014)\citenamefont {Wang},
  \citenamefont {Graf}, \citenamefont {Li}, \citenamefont {Wang},\ and\
  \citenamefont {Petrovic}}]{wang_anisotropic_2014}%
  \BibitemOpen
  \bibfield  {author} {\bibinfo {author} {\bibfnamefont {K.}~\bibnamefont
  {Wang}}, \bibinfo {author} {\bibfnamefont {D.}~\bibnamefont {Graf}}, \bibinfo
  {author} {\bibfnamefont {L.}~\bibnamefont {Li}}, \bibinfo {author}
  {\bibfnamefont {L.}~\bibnamefont {Wang}}, \ and\ \bibinfo {author}
  {\bibfnamefont {C.}~\bibnamefont {Petrovic}},\ }\href
  {http://www.nature.com/srep/2014/141205/srep07328/full/srep07328.html}
  {\bibfield  {journal} {\bibinfo  {journal} {Scientific Reports}\ }\textbf
  {\bibinfo {volume} {4}} (\bibinfo {year} {2014})}\BibitemShut {NoStop}%
\bibitem [{\citenamefont {Wang}\ \emph {et~al.}(2016)\citenamefont {Wang},
  \citenamefont {Li}, \citenamefont {Lu}, \citenamefont {Shen}, \citenamefont
  {Sheng}, \citenamefont {Feng}, \citenamefont {Zheng},\ and\ \citenamefont
  {Xu}}]{wang_topological_2016}%
  \BibitemOpen
  \bibfield  {author} {\bibinfo {author} {\bibfnamefont {Z.}~\bibnamefont
  {Wang}}, \bibinfo {author} {\bibfnamefont {Y.}~\bibnamefont {Li}}, \bibinfo
  {author} {\bibfnamefont {Y.}~\bibnamefont {Lu}}, \bibinfo {author}
  {\bibfnamefont {Z.}~\bibnamefont {Shen}}, \bibinfo {author} {\bibfnamefont
  {F.}~\bibnamefont {Sheng}}, \bibinfo {author} {\bibfnamefont
  {C.}~\bibnamefont {Feng}}, \bibinfo {author} {\bibfnamefont {Y.}~\bibnamefont
  {Zheng}}, \ and\ \bibinfo {author} {\bibfnamefont {Z.}~\bibnamefont {Xu}},\
  }\href {http://arxiv.org/abs/1603.01717} {\bibfield  {journal} {\bibinfo
  {journal} {arXiv:1603.01717 [cond-mat]}\ } (\bibinfo {year} {2016})},\
  \bibinfo {note} {arXiv: 1603.01717}\BibitemShut {NoStop}%
\bibitem [{\citenamefont {Ali}\ \emph {et~al.}(2014)\citenamefont {Ali},
  \citenamefont {Xiong}, \citenamefont {Flynn}, \citenamefont {Tao},
  \citenamefont {Gibson}, \citenamefont {Schoop}, \citenamefont {Liang},
  \citenamefont {Haldolaarachchige}, \citenamefont {Hirschberger},
  \citenamefont {Ong},\ and\ \citenamefont {Cava}}]{ali_large_2014}%
  \BibitemOpen
  \bibfield  {author} {\bibinfo {author} {\bibfnamefont {M.~N.}\ \bibnamefont
  {Ali}}, \bibinfo {author} {\bibfnamefont {J.}~\bibnamefont {Xiong}}, \bibinfo
  {author} {\bibfnamefont {S.}~\bibnamefont {Flynn}}, \bibinfo {author}
  {\bibfnamefont {J.}~\bibnamefont {Tao}}, \bibinfo {author} {\bibfnamefont
  {Q.~D.}\ \bibnamefont {Gibson}}, \bibinfo {author} {\bibfnamefont {L.~M.}\
  \bibnamefont {Schoop}}, \bibinfo {author} {\bibfnamefont {T.}~\bibnamefont
  {Liang}}, \bibinfo {author} {\bibfnamefont {N.}~\bibnamefont
  {Haldolaarachchige}}, \bibinfo {author} {\bibfnamefont {M.}~\bibnamefont
  {Hirschberger}}, \bibinfo {author} {\bibfnamefont {N.~P.}\ \bibnamefont
  {Ong}}, \ and\ \bibinfo {author} {\bibfnamefont {R.~J.}\ \bibnamefont
  {Cava}},\ }\href {\doibase 10.1038/nature13763} {\bibfield  {journal}
  {\bibinfo  {journal} {Nature}\ }\textbf {\bibinfo {volume} {514}},\ \bibinfo
  {pages} {205} (\bibinfo {year} {2014})}\BibitemShut {NoStop}%
\bibitem [{\citenamefont {Ali}\ \emph {et~al.}(2015)\citenamefont {Ali},
  \citenamefont {Schoop}, \citenamefont {Xiong}, \citenamefont {Flynn},
  \citenamefont {Gibson}, \citenamefont {Hirschberger}, \citenamefont {Ong},\
  and\ \citenamefont {Cava}}]{ali_correlation_2015}%
  \BibitemOpen
  \bibfield  {author} {\bibinfo {author} {\bibfnamefont {M.~N.}\ \bibnamefont
  {Ali}}, \bibinfo {author} {\bibfnamefont {L.}~\bibnamefont {Schoop}},
  \bibinfo {author} {\bibfnamefont {J.}~\bibnamefont {Xiong}}, \bibinfo
  {author} {\bibfnamefont {S.}~\bibnamefont {Flynn}}, \bibinfo {author}
  {\bibfnamefont {Q.}~\bibnamefont {Gibson}}, \bibinfo {author} {\bibfnamefont
  {M.}~\bibnamefont {Hirschberger}}, \bibinfo {author} {\bibfnamefont {N.~P.}\
  \bibnamefont {Ong}}, \ and\ \bibinfo {author} {\bibfnamefont {R.~J.}\
  \bibnamefont {Cava}},\ }\href {\doibase 10.1209/0295-5075/110/67002}
  {\bibfield  {journal} {\bibinfo  {journal} {EPL (Europhysics Letters)}\
  }\textbf {\bibinfo {volume} {110}},\ \bibinfo {pages} {67002} (\bibinfo
  {year} {2015})}\BibitemShut {NoStop}%
\bibitem [{\citenamefont {Tritt}\ \emph {et~al.}(1999)\citenamefont {Tritt},
  \citenamefont {Lowhorn}, \citenamefont {Littleton}, \citenamefont {Pope},
  \citenamefont {Feger},\ and\ \citenamefont {Kolis}}]{tritt_large_1999}%
  \BibitemOpen
  \bibfield  {author} {\bibinfo {author} {\bibfnamefont {T.~M.}\ \bibnamefont
  {Tritt}}, \bibinfo {author} {\bibfnamefont {N.~D.}\ \bibnamefont {Lowhorn}},
  \bibinfo {author} {\bibfnamefont {R.~T.}\ \bibnamefont {Littleton}}, \bibinfo
  {author} {\bibfnamefont {A.}~\bibnamefont {Pope}}, \bibinfo {author}
  {\bibfnamefont {C.~R.}\ \bibnamefont {Feger}}, \ and\ \bibinfo {author}
  {\bibfnamefont {J.~W.}\ \bibnamefont {Kolis}},\ }\href {\doibase
  10.1103/PhysRevB.60.7816} {\bibfield  {journal} {\bibinfo  {journal}
  {Physical Review B}\ }\textbf {\bibinfo {volume} {60}},\ \bibinfo {pages}
  {7816} (\bibinfo {year} {1999})}\BibitemShut {NoStop}%
\bibitem [{\citenamefont {Kang}\ \emph {et~al.}(2015)\citenamefont {Kang},
  \citenamefont {Zhou}, \citenamefont {Yi}, \citenamefont {Yang}, \citenamefont
  {Guo}, \citenamefont {Shi}, \citenamefont {Zhang}, \citenamefont {Wang},
  \citenamefont {Zhang}, \citenamefont {Jiang}, \citenamefont {Li},
  \citenamefont {Yang}, \citenamefont {Wu}, \citenamefont {Zhang},
  \citenamefont {Sun},\ and\ \citenamefont
  {Zhao}}]{kang_superconductivity_2015}%
  \BibitemOpen
  \bibfield  {author} {\bibinfo {author} {\bibfnamefont {D.}~\bibnamefont
  {Kang}}, \bibinfo {author} {\bibfnamefont {Y.}~\bibnamefont {Zhou}}, \bibinfo
  {author} {\bibfnamefont {W.}~\bibnamefont {Yi}}, \bibinfo {author}
  {\bibfnamefont {C.}~\bibnamefont {Yang}}, \bibinfo {author} {\bibfnamefont
  {J.}~\bibnamefont {Guo}}, \bibinfo {author} {\bibfnamefont {Y.}~\bibnamefont
  {Shi}}, \bibinfo {author} {\bibfnamefont {S.}~\bibnamefont {Zhang}}, \bibinfo
  {author} {\bibfnamefont {Z.}~\bibnamefont {Wang}}, \bibinfo {author}
  {\bibfnamefont {C.}~\bibnamefont {Zhang}}, \bibinfo {author} {\bibfnamefont
  {S.}~\bibnamefont {Jiang}}, \bibinfo {author} {\bibfnamefont
  {A.}~\bibnamefont {Li}}, \bibinfo {author} {\bibfnamefont {K.}~\bibnamefont
  {Yang}}, \bibinfo {author} {\bibfnamefont {Q.}~\bibnamefont {Wu}}, \bibinfo
  {author} {\bibfnamefont {G.}~\bibnamefont {Zhang}}, \bibinfo {author}
  {\bibfnamefont {L.}~\bibnamefont {Sun}}, \ and\ \bibinfo {author}
  {\bibfnamefont {Z.}~\bibnamefont {Zhao}},\ }\href {\doibase
  10.1038/ncomms8804} {\bibfield  {journal} {\bibinfo  {journal} {Nature
  Communications}\ }\textbf {\bibinfo {volume} {6}},\ \bibinfo {pages} {7804}
  (\bibinfo {year} {2015})}\BibitemShut {NoStop}%
\bibitem [{\citenamefont {Pan}\ \emph {et~al.}(2015)\citenamefont {Pan},
  \citenamefont {Chen}, \citenamefont {Liu}, \citenamefont {Feng},
  \citenamefont {Wei}, \citenamefont {Zhou}, \citenamefont {Chi}, \citenamefont
  {Pi}, \citenamefont {Yen}, \citenamefont {Song}, \citenamefont {Wan},
  \citenamefont {Yang}, \citenamefont {Wang}, \citenamefont {Wang},\ and\
  \citenamefont {Zhang}}]{pan_pressure-driven_2015}%
  \BibitemOpen
  \bibfield  {author} {\bibinfo {author} {\bibfnamefont {X.-C.}\ \bibnamefont
  {Pan}}, \bibinfo {author} {\bibfnamefont {X.}~\bibnamefont {Chen}}, \bibinfo
  {author} {\bibfnamefont {H.}~\bibnamefont {Liu}}, \bibinfo {author}
  {\bibfnamefont {Y.}~\bibnamefont {Feng}}, \bibinfo {author} {\bibfnamefont
  {Z.}~\bibnamefont {Wei}}, \bibinfo {author} {\bibfnamefont {Y.}~\bibnamefont
  {Zhou}}, \bibinfo {author} {\bibfnamefont {Z.}~\bibnamefont {Chi}}, \bibinfo
  {author} {\bibfnamefont {L.}~\bibnamefont {Pi}}, \bibinfo {author}
  {\bibfnamefont {F.}~\bibnamefont {Yen}}, \bibinfo {author} {\bibfnamefont
  {F.}~\bibnamefont {Song}}, \bibinfo {author} {\bibfnamefont {X.}~\bibnamefont
  {Wan}}, \bibinfo {author} {\bibfnamefont {Z.}~\bibnamefont {Yang}}, \bibinfo
  {author} {\bibfnamefont {B.}~\bibnamefont {Wang}}, \bibinfo {author}
  {\bibfnamefont {G.}~\bibnamefont {Wang}}, \ and\ \bibinfo {author}
  {\bibfnamefont {Y.}~\bibnamefont {Zhang}},\ }\href {\doibase
  10.1038/ncomms8805} {\bibfield  {journal} {\bibinfo  {journal} {Nature
  Communications}\ }\textbf {\bibinfo {volume} {6}},\ \bibinfo {pages} {7805}
  (\bibinfo {year} {2015})}\BibitemShut {NoStop}%
\bibitem [{\citenamefont {Zhou}\ \emph {et~al.}(2016)\citenamefont {Zhou},
  \citenamefont {Wu}, \citenamefont {Ning}, \citenamefont {Li}, \citenamefont
  {Du}, \citenamefont {Chen}, \citenamefont {Zhang}, \citenamefont {Chi},
  \citenamefont {Wang}, \citenamefont {Zhu}, \citenamefont {Lu}, \citenamefont
  {Ji}, \citenamefont {Wan}, \citenamefont {Yang}, \citenamefont {Sun},
  \citenamefont {Yang}, \citenamefont {Tian}, \citenamefont {Zhang},\ and\
  \citenamefont {Mao}}]{zhou_pressure-induced_2016}%
  \BibitemOpen
  \bibfield  {author} {\bibinfo {author} {\bibfnamefont {Y.}~\bibnamefont
  {Zhou}}, \bibinfo {author} {\bibfnamefont {J.}~\bibnamefont {Wu}}, \bibinfo
  {author} {\bibfnamefont {W.}~\bibnamefont {Ning}}, \bibinfo {author}
  {\bibfnamefont {N.}~\bibnamefont {Li}}, \bibinfo {author} {\bibfnamefont
  {Y.}~\bibnamefont {Du}}, \bibinfo {author} {\bibfnamefont {X.}~\bibnamefont
  {Chen}}, \bibinfo {author} {\bibfnamefont {R.}~\bibnamefont {Zhang}},
  \bibinfo {author} {\bibfnamefont {Z.}~\bibnamefont {Chi}}, \bibinfo {author}
  {\bibfnamefont {X.}~\bibnamefont {Wang}}, \bibinfo {author} {\bibfnamefont
  {X.}~\bibnamefont {Zhu}}, \bibinfo {author} {\bibfnamefont {P.}~\bibnamefont
  {Lu}}, \bibinfo {author} {\bibfnamefont {C.}~\bibnamefont {Ji}}, \bibinfo
  {author} {\bibfnamefont {X.}~\bibnamefont {Wan}}, \bibinfo {author}
  {\bibfnamefont {Z.}~\bibnamefont {Yang}}, \bibinfo {author} {\bibfnamefont
  {J.}~\bibnamefont {Sun}}, \bibinfo {author} {\bibfnamefont {W.}~\bibnamefont
  {Yang}}, \bibinfo {author} {\bibfnamefont {M.}~\bibnamefont {Tian}}, \bibinfo
  {author} {\bibfnamefont {Y.}~\bibnamefont {Zhang}}, \ and\ \bibinfo {author}
  {\bibfnamefont {H.-k.}\ \bibnamefont {Mao}},\ }\href {\doibase
  10.1073/pnas.1601262113} {\bibfield  {journal} {\bibinfo  {journal}
  {Proceedings of the National Academy of Sciences}\ }\textbf {\bibinfo
  {volume} {113}},\ \bibinfo {pages} {2904} (\bibinfo {year}
  {2016})}\BibitemShut {NoStop}%
\bibitem [{\citenamefont {Qi}\ \emph {et~al.}(2016{\natexlab{a}})\citenamefont
  {Qi}, \citenamefont {Shi}, \citenamefont {Naumov}, \citenamefont {Kumar},
  \citenamefont {Schnelle}, \citenamefont {Barkalov}, \citenamefont {Shekhar},
  \citenamefont {Borrmann}, \citenamefont {Felser}, \citenamefont {Yan},\ and\
  \citenamefont {Medvedev}}]{qi_pressure-driven_2016}%
  \BibitemOpen
  \bibfield  {author} {\bibinfo {author} {\bibfnamefont {Y.}~\bibnamefont
  {Qi}}, \bibinfo {author} {\bibfnamefont {W.}~\bibnamefont {Shi}}, \bibinfo
  {author} {\bibfnamefont {P.~G.}\ \bibnamefont {Naumov}}, \bibinfo {author}
  {\bibfnamefont {N.}~\bibnamefont {Kumar}}, \bibinfo {author} {\bibfnamefont
  {W.}~\bibnamefont {Schnelle}}, \bibinfo {author} {\bibfnamefont
  {O.}~\bibnamefont {Barkalov}}, \bibinfo {author} {\bibfnamefont
  {C.}~\bibnamefont {Shekhar}}, \bibinfo {author} {\bibfnamefont
  {H.}~\bibnamefont {Borrmann}}, \bibinfo {author} {\bibfnamefont
  {C.}~\bibnamefont {Felser}}, \bibinfo {author} {\bibfnamefont
  {B.}~\bibnamefont {Yan}}, \ and\ \bibinfo {author} {\bibfnamefont {S.~A.}\
  \bibnamefont {Medvedev}},\ }\href {http://arxiv.org/abs/1602.08616}
  {\bibfield  {journal} {\bibinfo  {journal} {arXiv:1602.08616 [cond-mat]}\ }
  (\bibinfo {year} {2016}{\natexlab{a}})},\ \bibinfo {note} {arXiv:
  1602.08616}\BibitemShut {NoStop}%
\bibitem [{\citenamefont {Qi}\ \emph {et~al.}(2016{\natexlab{b}})\citenamefont
  {Qi}, \citenamefont {Naumov}, \citenamefont {Ali}, \citenamefont {Rajamathi},
  \citenamefont {Schnelle}, \citenamefont {Barkalov}, \citenamefont {Hanfland},
  \citenamefont {Wu}, \citenamefont {Shekhar}, \citenamefont {Sun},
  \citenamefont {Süß}, \citenamefont {Schmidt}, \citenamefont {Schwarz},
  \citenamefont {Pippel}, \citenamefont {Werner}, \citenamefont {Hillebrand},
  \citenamefont {Förster}, \citenamefont {Kampert}, \citenamefont {Parkin},
  \citenamefont {Cava}, \citenamefont {Felser}, \citenamefont {Yan},\ and\
  \citenamefont {Medvedev}}]{qi_superconductivity_2016}%
  \BibitemOpen
  \bibfield  {author} {\bibinfo {author} {\bibfnamefont {Y.}~\bibnamefont
  {Qi}}, \bibinfo {author} {\bibfnamefont {P.~G.}\ \bibnamefont {Naumov}},
  \bibinfo {author} {\bibfnamefont {M.~N.}\ \bibnamefont {Ali}}, \bibinfo
  {author} {\bibfnamefont {C.~R.}\ \bibnamefont {Rajamathi}}, \bibinfo {author}
  {\bibfnamefont {W.}~\bibnamefont {Schnelle}}, \bibinfo {author}
  {\bibfnamefont {O.}~\bibnamefont {Barkalov}}, \bibinfo {author}
  {\bibfnamefont {M.}~\bibnamefont {Hanfland}}, \bibinfo {author}
  {\bibfnamefont {S.-C.}\ \bibnamefont {Wu}}, \bibinfo {author} {\bibfnamefont
  {C.}~\bibnamefont {Shekhar}}, \bibinfo {author} {\bibfnamefont
  {Y.}~\bibnamefont {Sun}}, \bibinfo {author} {\bibfnamefont {V.}~\bibnamefont
  {Süß}}, \bibinfo {author} {\bibfnamefont {M.}~\bibnamefont {Schmidt}},
  \bibinfo {author} {\bibfnamefont {U.}~\bibnamefont {Schwarz}}, \bibinfo
  {author} {\bibfnamefont {E.}~\bibnamefont {Pippel}}, \bibinfo {author}
  {\bibfnamefont {P.}~\bibnamefont {Werner}}, \bibinfo {author} {\bibfnamefont
  {R.}~\bibnamefont {Hillebrand}}, \bibinfo {author} {\bibfnamefont
  {T.}~\bibnamefont {Förster}}, \bibinfo {author} {\bibfnamefont
  {E.}~\bibnamefont {Kampert}}, \bibinfo {author} {\bibfnamefont
  {S.}~\bibnamefont {Parkin}}, \bibinfo {author} {\bibfnamefont {R.~J.}\
  \bibnamefont {Cava}}, \bibinfo {author} {\bibfnamefont {C.}~\bibnamefont
  {Felser}}, \bibinfo {author} {\bibfnamefont {B.}~\bibnamefont {Yan}}, \ and\
  \bibinfo {author} {\bibfnamefont {S.~A.}\ \bibnamefont {Medvedev}},\ }\href
  {\doibase 10.1038/ncomms11038} {\bibfield  {journal} {\bibinfo  {journal}
  {Nature Communications}\ }\textbf {\bibinfo {volume} {7}},\ \bibinfo {pages}
  {11038} (\bibinfo {year} {2016}{\natexlab{b}})}\BibitemShut {NoStop}%
\bibitem [{\citenamefont {Zeng}\ \emph {et~al.}(2015)\citenamefont {Zeng},
  \citenamefont {Fang}, \citenamefont {Chang}, \citenamefont {Chen},
  \citenamefont {Hsieh}, \citenamefont {Bansil}, \citenamefont {Lin},\ and\
  \citenamefont {Fu}}]{zeng_topological_2015}%
  \BibitemOpen
  \bibfield  {author} {\bibinfo {author} {\bibfnamefont {M.}~\bibnamefont
  {Zeng}}, \bibinfo {author} {\bibfnamefont {C.}~\bibnamefont {Fang}}, \bibinfo
  {author} {\bibfnamefont {G.}~\bibnamefont {Chang}}, \bibinfo {author}
  {\bibfnamefont {Y.-A.}\ \bibnamefont {Chen}}, \bibinfo {author}
  {\bibfnamefont {T.}~\bibnamefont {Hsieh}}, \bibinfo {author} {\bibfnamefont
  {A.}~\bibnamefont {Bansil}}, \bibinfo {author} {\bibfnamefont
  {H.}~\bibnamefont {Lin}}, \ and\ \bibinfo {author} {\bibfnamefont
  {L.}~\bibnamefont {Fu}},\ }\href@noop {} {\bibfield  {journal} {\bibinfo
  {journal} {arXiv:1504.03492 [cond-mat]}\ } (\bibinfo {year} {2015})},\
  \bibinfo {note} {arXiv: 1504.03492}\BibitemShut {NoStop}%
\bibitem [{\citenamefont {Tafti}\ \emph
  {et~al.}(2016{\natexlab{a}})\citenamefont {Tafti}, \citenamefont {Gibson},
  \citenamefont {Kushwaha}, \citenamefont {Krizan}, \citenamefont
  {Haldolaarachchige},\ and\ \citenamefont
  {Cava}}]{tafti_temperaturefield_2016}%
  \BibitemOpen
  \bibfield  {author} {\bibinfo {author} {\bibfnamefont {F.~F.}\ \bibnamefont
  {Tafti}}, \bibinfo {author} {\bibfnamefont {Q.}~\bibnamefont {Gibson}},
  \bibinfo {author} {\bibfnamefont {S.}~\bibnamefont {Kushwaha}}, \bibinfo
  {author} {\bibfnamefont {J.~W.}\ \bibnamefont {Krizan}}, \bibinfo {author}
  {\bibfnamefont {N.}~\bibnamefont {Haldolaarachchige}}, \ and\ \bibinfo
  {author} {\bibfnamefont {R.~J.}\ \bibnamefont {Cava}},\ }\href {\doibase
  10.1073/pnas.1607319113} {\bibfield  {journal} {\bibinfo  {journal}
  {Proceedings of the National Academy of Sciences}\ }\textbf {\bibinfo
  {volume} {113}},\ \bibinfo {pages} {E3475} (\bibinfo {year}
  {2016}{\natexlab{a}})}\BibitemShut {NoStop}%
\bibitem [{\citenamefont {Tafti}\ \emph
  {et~al.}(2016{\natexlab{b}})\citenamefont {Tafti}, \citenamefont {Gibson},
  \citenamefont {Kushwaha}, \citenamefont {Haldolaarachchige},\ and\
  \citenamefont {Cava}}]{tafti_resistivity_2016}%
  \BibitemOpen
  \bibfield  {author} {\bibinfo {author} {\bibfnamefont {F.~F.}\ \bibnamefont
  {Tafti}}, \bibinfo {author} {\bibfnamefont {Q.~D.}\ \bibnamefont {Gibson}},
  \bibinfo {author} {\bibfnamefont {S.~K.}\ \bibnamefont {Kushwaha}}, \bibinfo
  {author} {\bibfnamefont {N.}~\bibnamefont {Haldolaarachchige}}, \ and\
  \bibinfo {author} {\bibfnamefont {R.~J.}\ \bibnamefont {Cava}},\ }\href
  {\doibase 10.1038/nphys3581} {\bibfield  {journal} {\bibinfo  {journal}
  {Nature Physics}\ }\textbf {\bibinfo {volume} {12}},\ \bibinfo {pages} {272}
  (\bibinfo {year} {2016}{\natexlab{b}})}\BibitemShut {NoStop}%
\bibitem [{\citenamefont {Zeng}\ \emph {et~al.}(2016)\citenamefont {Zeng},
  \citenamefont {Lou}, \citenamefont {Wu}, \citenamefont {Guo}, \citenamefont
  {Kong}, \citenamefont {Zhong}, \citenamefont {Ma}, \citenamefont {Fu},
  \citenamefont {Richard}, \citenamefont {Wang}, \citenamefont {Liu},
  \citenamefont {Lu}, \citenamefont {Sun}, \citenamefont {Wang}, \citenamefont
  {Wang}, \citenamefont {Shi}, \citenamefont {Lei}, \citenamefont {Liu},
  \citenamefont {Wang}, \citenamefont {Qian}, \citenamefont {Luo},\ and\
  \citenamefont {Ding}}]{zeng_compensated_2016}%
  \BibitemOpen
  \bibfield  {author} {\bibinfo {author} {\bibfnamefont {L.-K.}\ \bibnamefont
  {Zeng}}, \bibinfo {author} {\bibfnamefont {R.}~\bibnamefont {Lou}}, \bibinfo
  {author} {\bibfnamefont {D.-S.}\ \bibnamefont {Wu}}, \bibinfo {author}
  {\bibfnamefont {P.-J.}\ \bibnamefont {Guo}}, \bibinfo {author} {\bibfnamefont
  {L.-Y.}\ \bibnamefont {Kong}}, \bibinfo {author} {\bibfnamefont {Y.-G.}\
  \bibnamefont {Zhong}}, \bibinfo {author} {\bibfnamefont {J.-Z.}\ \bibnamefont
  {Ma}}, \bibinfo {author} {\bibfnamefont {B.-B.}\ \bibnamefont {Fu}}, \bibinfo
  {author} {\bibfnamefont {P.}~\bibnamefont {Richard}}, \bibinfo {author}
  {\bibfnamefont {P.}~\bibnamefont {Wang}}, \bibinfo {author} {\bibfnamefont
  {G.~T.}\ \bibnamefont {Liu}}, \bibinfo {author} {\bibfnamefont
  {L.}~\bibnamefont {Lu}}, \bibinfo {author} {\bibfnamefont {S.-S.}\
  \bibnamefont {Sun}}, \bibinfo {author} {\bibfnamefont {Q.}~\bibnamefont
  {Wang}}, \bibinfo {author} {\bibfnamefont {L.}~\bibnamefont {Wang}}, \bibinfo
  {author} {\bibfnamefont {Y.-G.}\ \bibnamefont {Shi}}, \bibinfo {author}
  {\bibfnamefont {H.-C.}\ \bibnamefont {Lei}}, \bibinfo {author} {\bibfnamefont
  {K.}~\bibnamefont {Liu}}, \bibinfo {author} {\bibfnamefont {S.-C.}\
  \bibnamefont {Wang}}, \bibinfo {author} {\bibfnamefont {T.}~\bibnamefont
  {Qian}}, \bibinfo {author} {\bibfnamefont {J.-L.}\ \bibnamefont {Luo}}, \
  and\ \bibinfo {author} {\bibfnamefont {H.}~\bibnamefont {Ding}},\ }\href
  {http://arxiv.org/abs/1604.08142} {\bibfield  {journal} {\bibinfo  {journal}
  {arXiv:1604.08142 [cond-mat]}\ } (\bibinfo {year} {2016})},\ \bibinfo {note}
  {arXiv: 1604.08142}\BibitemShut {NoStop}%
\bibitem [{\citenamefont {Wu}\ \emph {et~al.}(2016)\citenamefont {Wu},
  \citenamefont {Kong}, \citenamefont {Wang}, \citenamefont {Johnson},
  \citenamefont {Mou}, \citenamefont {Huang}, \citenamefont {Schrunk},
  \citenamefont {Bud'ko}, \citenamefont {Canfield},\ and\ \citenamefont
  {Kaminski}}]{wu_unusual_2016}%
  \BibitemOpen
  \bibfield  {author} {\bibinfo {author} {\bibfnamefont {Y.}~\bibnamefont
  {Wu}}, \bibinfo {author} {\bibfnamefont {T.}~\bibnamefont {Kong}}, \bibinfo
  {author} {\bibfnamefont {L.-L.}\ \bibnamefont {Wang}}, \bibinfo {author}
  {\bibfnamefont {D.~D.}\ \bibnamefont {Johnson}}, \bibinfo {author}
  {\bibfnamefont {D.}~\bibnamefont {Mou}}, \bibinfo {author} {\bibfnamefont
  {L.}~\bibnamefont {Huang}}, \bibinfo {author} {\bibfnamefont
  {B.}~\bibnamefont {Schrunk}}, \bibinfo {author} {\bibfnamefont {S.~L.}\
  \bibnamefont {Bud'ko}}, \bibinfo {author} {\bibfnamefont {P.~C.}\
  \bibnamefont {Canfield}}, \ and\ \bibinfo {author} {\bibfnamefont
  {A.}~\bibnamefont {Kaminski}},\ }\href {http://arxiv.org/abs/1604.08945}
  {\bibfield  {journal} {\bibinfo  {journal} {arXiv:1604.08945 [cond-mat]}\ }
  (\bibinfo {year} {2016})},\ \bibinfo {note} {arXiv: 1604.08945}\BibitemShut
  {NoStop}%
\bibitem [{\citenamefont {Nayak}\ \emph {et~al.}(2016)\citenamefont {Nayak},
  \citenamefont {Wu}, \citenamefont {Kumar}, \citenamefont {Shekhar},
  \citenamefont {Singh}, \citenamefont {Fink}, \citenamefont {Rienks},
  \citenamefont {Fecher}, \citenamefont {Parkin}, \citenamefont {Yan},\ and\
  \citenamefont {Felser}}]{nayak_multiple_2016}%
  \BibitemOpen
  \bibfield  {author} {\bibinfo {author} {\bibfnamefont {J.}~\bibnamefont
  {Nayak}}, \bibinfo {author} {\bibfnamefont {S.-C.}\ \bibnamefont {Wu}},
  \bibinfo {author} {\bibfnamefont {N.}~\bibnamefont {Kumar}}, \bibinfo
  {author} {\bibfnamefont {C.}~\bibnamefont {Shekhar}}, \bibinfo {author}
  {\bibfnamefont {S.}~\bibnamefont {Singh}}, \bibinfo {author} {\bibfnamefont
  {J.}~\bibnamefont {Fink}}, \bibinfo {author} {\bibfnamefont {E.~E.~D.}\
  \bibnamefont {Rienks}}, \bibinfo {author} {\bibfnamefont {G.~H.}\
  \bibnamefont {Fecher}}, \bibinfo {author} {\bibfnamefont {S.~S.~P.}\
  \bibnamefont {Parkin}}, \bibinfo {author} {\bibfnamefont {B.}~\bibnamefont
  {Yan}}, \ and\ \bibinfo {author} {\bibfnamefont {C.}~\bibnamefont {Felser}},\
  }\href {http://arxiv.org/abs/1605.06997} {\bibfield  {journal} {\bibinfo
  {journal} {arXiv:1605.06997 [cond-mat]}\ } (\bibinfo {year} {2016})},\
  \bibinfo {note} {arXiv: 1605.06997}\BibitemShut {NoStop}%
\bibitem [{\citenamefont {Sun}\ \emph {et~al.}(2016)\citenamefont {Sun},
  \citenamefont {Wang}, \citenamefont {Guo}, \citenamefont {Liu},\ and\
  \citenamefont {Lei}}]{sun_large_2016}%
  \BibitemOpen
  \bibfield  {author} {\bibinfo {author} {\bibfnamefont {S.}~\bibnamefont
  {Sun}}, \bibinfo {author} {\bibfnamefont {Q.}~\bibnamefont {Wang}}, \bibinfo
  {author} {\bibfnamefont {P.-J.}\ \bibnamefont {Guo}}, \bibinfo {author}
  {\bibfnamefont {K.}~\bibnamefont {Liu}}, \ and\ \bibinfo {author}
  {\bibfnamefont {H.}~\bibnamefont {Lei}},\ }\href
  {http://arxiv.org/abs/1601.04618} {\bibfield  {journal} {\bibinfo  {journal}
  {arXiv:1601.04618 [cond-mat]}\ } (\bibinfo {year} {2016})},\ \bibinfo {note}
  {arXiv: 1601.04618}\BibitemShut {NoStop}%
\bibitem [{\citenamefont {Eiling}\ and\ \citenamefont
  {Schilling}(1981)}]{eiling_pressure_1981}%
  \BibitemOpen
  \bibfield  {author} {\bibinfo {author} {\bibfnamefont {A.}~\bibnamefont
  {Eiling}}\ and\ \bibinfo {author} {\bibfnamefont {J.~S.}\ \bibnamefont
  {Schilling}},\ }\href {\doibase 10.1088/0305-4608/11/3/010} {\bibfield
  {journal} {\bibinfo  {journal} {Journal of Physics F: Metal Physics}\
  }\textbf {\bibinfo {volume} {11}},\ \bibinfo {pages} {623} (\bibinfo {year}
  {1981})}\BibitemShut {NoStop}%
\bibitem [{\citenamefont {Weir}\ \emph {et~al.}(2000)\citenamefont {Weir},
  \citenamefont {Akella}, \citenamefont {Aracne-Ruddle}, \citenamefont
  {Vohra},\ and\ \citenamefont {Catledge}}]{weir_epitaxial_2000}%
  \BibitemOpen
  \bibfield  {author} {\bibinfo {author} {\bibfnamefont {S.~T.}\ \bibnamefont
  {Weir}}, \bibinfo {author} {\bibfnamefont {J.}~\bibnamefont {Akella}},
  \bibinfo {author} {\bibfnamefont {C.}~\bibnamefont {Aracne-Ruddle}}, \bibinfo
  {author} {\bibfnamefont {Y.~K.}\ \bibnamefont {Vohra}}, \ and\ \bibinfo
  {author} {\bibfnamefont {S.~A.}\ \bibnamefont {Catledge}},\ }\href {\doibase
  10.1063/1.1326838} {\bibfield  {journal} {\bibinfo  {journal} {Applied
  Physics Letters}\ }\textbf {\bibinfo {volume} {77}},\ \bibinfo {pages} {3400}
  (\bibinfo {year} {2000})}\BibitemShut {NoStop}%
\bibitem [{\citenamefont {Piermarini}\ \emph {et~al.}(1975)\citenamefont
  {Piermarini}, \citenamefont {Block}, \citenamefont {Barnett},\ and\
  \citenamefont {Forman}}]{piermarini_calibration_1975}%
  \BibitemOpen
  \bibfield  {author} {\bibinfo {author} {\bibfnamefont {G.~J.}\ \bibnamefont
  {Piermarini}}, \bibinfo {author} {\bibfnamefont {S.}~\bibnamefont {Block}},
  \bibinfo {author} {\bibfnamefont {J.~D.}\ \bibnamefont {Barnett}}, \ and\
  \bibinfo {author} {\bibfnamefont {R.~A.}\ \bibnamefont {Forman}},\ }\href
  {\doibase 10.1063/1.321957} {\bibfield  {journal} {\bibinfo  {journal}
  {Journal of Applied Physics}\ }\textbf {\bibinfo {volume} {46}},\ \bibinfo
  {pages} {2774} (\bibinfo {year} {1975})}\BibitemShut {NoStop}%
\bibitem [{\citenamefont {Hammersley}\ \emph {et~al.}(1996)\citenamefont
  {Hammersley}, \citenamefont {Svensson}, \citenamefont {Hanfland},
  \citenamefont {Fitch},\ and\ \citenamefont
  {Hausermann}}]{hammersley_two-dimensional_1996}%
  \BibitemOpen
  \bibfield  {author} {\bibinfo {author} {\bibfnamefont {A.~P.}\ \bibnamefont
  {Hammersley}}, \bibinfo {author} {\bibfnamefont {S.~O.}\ \bibnamefont
  {Svensson}}, \bibinfo {author} {\bibfnamefont {M.}~\bibnamefont {Hanfland}},
  \bibinfo {author} {\bibfnamefont {A.~N.}\ \bibnamefont {Fitch}}, \ and\
  \bibinfo {author} {\bibfnamefont {D.}~\bibnamefont {Hausermann}},\ }\href
  {\doibase 10.1080/08957959608201408} {\bibfield  {journal} {\bibinfo
  {journal} {High Pressure Research}\ }\textbf {\bibinfo {volume} {14}},\
  \bibinfo {pages} {235} (\bibinfo {year} {1996})}\BibitemShut {NoStop}%
\bibitem [{\citenamefont {Toby}(2001)}]{toby_xpgui_2001}%
  \BibitemOpen
  \bibfield  {author} {\bibinfo {author} {\bibfnamefont {B.~H.}\ \bibnamefont
  {Toby}},\ }\href {\doibase 10.1107/S0021889801002242} {\bibfield  {journal}
  {\bibinfo  {journal} {Journal of Applied Crystallography}\ }\textbf {\bibinfo
  {volume} {34}},\ \bibinfo {pages} {210} (\bibinfo {year} {2001})}\BibitemShut
  {NoStop}%
\bibitem [{\citenamefont {Blaha}\ \emph {et~al.}(2001)\citenamefont {Blaha},
  \citenamefont {Schwarz}, \citenamefont {Madsen}, \citenamefont {Kvasnicka},\
  and\ \citenamefont {Luitz}}]{blaha_wien2k_2001}%
  \BibitemOpen
  \bibfield  {author} {\bibinfo {author} {\bibfnamefont {P.}~\bibnamefont
  {Blaha}}, \bibinfo {author} {\bibfnamefont {K.}~\bibnamefont {Schwarz}},
  \bibinfo {author} {\bibfnamefont {G.}~\bibnamefont {Madsen}}, \bibinfo
  {author} {\bibfnamefont {D.}~\bibnamefont {Kvasnicka}}, \ and\ \bibinfo
  {author} {\bibfnamefont {J.}~\bibnamefont {Luitz}},\ }\href@noop {} {\emph
  {\bibinfo {title} {WIEN2K, An Augmented Plane Wave + Local Orbitals Program
  for Calculating Crystal Properties}}}\ (\bibinfo  {publisher} {Karlheinz
  Schwarz, Techn. Universität Wien, Austria},\ \bibinfo {address} {Wien,
  Austria},\ \bibinfo {year} {2001})\BibitemShut {NoStop}%
\bibitem [{\citenamefont {Vaitheeswaran}\ \emph {et~al.}(2002)\citenamefont
  {Vaitheeswaran}, \citenamefont {Kanchana},\ and\ \citenamefont
  {Rajagopalan}}]{vaitheeswaran_electronic_2002}%
  \BibitemOpen
  \bibfield  {author} {\bibinfo {author} {\bibfnamefont {G.}~\bibnamefont
  {Vaitheeswaran}}, \bibinfo {author} {\bibfnamefont {V.}~\bibnamefont
  {Kanchana}}, \ and\ \bibinfo {author} {\bibfnamefont {M.}~\bibnamefont
  {Rajagopalan}},\ }\href {\doibase 10.1016/S0921-4526(01)01460-0} {\bibfield
  {journal} {\bibinfo  {journal} {Physica B: Condensed Matter}\ }\textbf
  {\bibinfo {volume} {315}},\ \bibinfo {pages} {64} (\bibinfo {year}
  {2002})}\BibitemShut {NoStop}%
\bibitem [{\citenamefont {Charifi}\ \emph {et~al.}(2008)\citenamefont
  {Charifi}, \citenamefont {Reshak},\ and\ \citenamefont
  {Baaziz}}]{charifi_phase_2008}%
  \BibitemOpen
  \bibfield  {author} {\bibinfo {author} {\bibfnamefont {Z.}~\bibnamefont
  {Charifi}}, \bibinfo {author} {\bibfnamefont {A.~H.}\ \bibnamefont {Reshak}},
  \ and\ \bibinfo {author} {\bibfnamefont {H.}~\bibnamefont {Baaziz}},\ }\href
  {\doibase 10.1016/j.ssc.2008.07.038} {\bibfield  {journal} {\bibinfo
  {journal} {Solid State Communications}\ }\textbf {\bibinfo {volume} {148}},\
  \bibinfo {pages} {139} (\bibinfo {year} {2008})}\BibitemShut {NoStop}%
\bibitem [{\citenamefont {Cui}\ \emph {et~al.}(2009)\citenamefont {Cui},
  \citenamefont {Feng}, \citenamefont {Hu}, \citenamefont {Feng},\ and\
  \citenamefont {Liu}}]{cui_first-principles_2009}%
  \BibitemOpen
  \bibfield  {author} {\bibinfo {author} {\bibfnamefont {S.}~\bibnamefont
  {Cui}}, \bibinfo {author} {\bibfnamefont {W.}~\bibnamefont {Feng}}, \bibinfo
  {author} {\bibfnamefont {H.}~\bibnamefont {Hu}}, \bibinfo {author}
  {\bibfnamefont {Z.}~\bibnamefont {Feng}}, \ and\ \bibinfo {author}
  {\bibfnamefont {H.}~\bibnamefont {Liu}},\ }\href {\doibase
  10.1016/j.ssc.2009.04.012} {\bibfield  {journal} {\bibinfo  {journal} {Solid
  State Communications}\ }\textbf {\bibinfo {volume} {149}},\ \bibinfo {pages}
  {996} (\bibinfo {year} {2009})}\BibitemShut {NoStop}%
\bibitem [{\citenamefont {Murnaghan}(1937)}]{murnaghan_finite_1937}%
  \BibitemOpen
  \bibfield  {author} {\bibinfo {author} {\bibfnamefont {F.~D.}\ \bibnamefont
  {Murnaghan}},\ }\href {\doibase 10.2307/2371405} {\bibfield  {journal}
  {\bibinfo  {journal} {American Journal of Mathematics}\ }\textbf {\bibinfo
  {volume} {59}},\ \bibinfo {pages} {235} (\bibinfo {year} {1937})}\BibitemShut
  {NoStop}%
\bibitem [{\citenamefont {Birch}(1947)}]{birch_finite_1947}%
  \BibitemOpen
  \bibfield  {author} {\bibinfo {author} {\bibfnamefont {F.}~\bibnamefont
  {Birch}},\ }\href {\doibase 10.1103/PhysRev.71.809} {\bibfield  {journal}
  {\bibinfo  {journal} {Physical Review}\ }\textbf {\bibinfo {volume} {71}},\
  \bibinfo {pages} {809} (\bibinfo {year} {1947})}\BibitemShut {NoStop}%
\bibitem [{\citenamefont {Fu}\ \emph {et~al.}(2007)\citenamefont {Fu},
  \citenamefont {Kane},\ and\ \citenamefont {Mele}}]{fu_topological_2007}%
  \BibitemOpen
  \bibfield  {author} {\bibinfo {author} {\bibfnamefont {L.}~\bibnamefont
  {Fu}}, \bibinfo {author} {\bibfnamefont {C.~L.}\ \bibnamefont {Kane}}, \ and\
  \bibinfo {author} {\bibfnamefont {E.~J.}\ \bibnamefont {Mele}},\ }\href
  {\doibase 10.1103/PhysRevLett.98.106803} {\bibfield  {journal} {\bibinfo
  {journal} {Physical Review Letters}\ }\textbf {\bibinfo {volume} {98}},\
  \bibinfo {pages} {106803} (\bibinfo {year} {2007})}\BibitemShut {NoStop}%
\bibitem [{\citenamefont {Il'ina}\ \emph {et~al.}(1972)\citenamefont {Il'ina},
  \citenamefont {Itskevich},\ and\ \citenamefont {Dizhur}}]{ilina_1972}%
  \BibitemOpen
  \bibfield  {author} {\bibinfo {author} {\bibfnamefont {M.}~\bibnamefont
  {Il'ina}}, \bibinfo {author} {\bibfnamefont {E.}~\bibnamefont {Itskevich}}, \
  and\ \bibinfo {author} {\bibfnamefont {E.}~\bibnamefont {Dizhur}},\
  }\href@noop {} {\bibfield  {journal} {\bibinfo  {journal} {Low Temperature
  Physics}\ }\textbf {\bibinfo {volume} {34}},\ \bibinfo {pages} {1263}
  (\bibinfo {year} {1972})}\BibitemShut {NoStop}%
\bibitem [{\citenamefont {Baring}\ \emph {et~al.}(2011)\citenamefont {Baring},
  \citenamefont {Silva},\ and\ \citenamefont {Kopelevich}}]{baring_local_2011}%
  \BibitemOpen
  \bibfield  {author} {\bibinfo {author} {\bibfnamefont {L.~A.}\ \bibnamefont
  {Baring}}, \bibinfo {author} {\bibfnamefont {R.~R.~d.}\ \bibnamefont
  {Silva}}, \ and\ \bibinfo {author} {\bibfnamefont {Y.}~\bibnamefont
  {Kopelevich}},\ }\href {\doibase 10.1063/1.3671591} {\bibfield  {journal}
  {\bibinfo  {journal} {Low Temperature Physics}\ }\textbf {\bibinfo {volume}
  {37}},\ \bibinfo {pages} {889} (\bibinfo {year} {2011})}\BibitemShut
  {NoStop}%
\bibitem [{\citenamefont {Zhu}\ \emph {et~al.}(2008)\citenamefont {Zhu},
  \citenamefont {Yang}, \citenamefont {Fang}, \citenamefont {Mu},\ and\
  \citenamefont {Wen}}]{zhu_upper_2008}%
  \BibitemOpen
  \bibfield  {author} {\bibinfo {author} {\bibfnamefont {X.}~\bibnamefont
  {Zhu}}, \bibinfo {author} {\bibfnamefont {H.}~\bibnamefont {Yang}}, \bibinfo
  {author} {\bibfnamefont {L.}~\bibnamefont {Fang}}, \bibinfo {author}
  {\bibfnamefont {G.}~\bibnamefont {Mu}}, \ and\ \bibinfo {author}
  {\bibfnamefont {H.-H.}\ \bibnamefont {Wen}},\ }\href {\doibase
  10.1088/0953-2048/21/10/105001} {\bibfield  {journal} {\bibinfo  {journal}
  {Superconductor Science and Technology}\ }\textbf {\bibinfo {volume} {21}},\
  \bibinfo {pages} {105001} (\bibinfo {year} {2008})}\BibitemShut {NoStop}%
\bibitem [{\citenamefont {Fang}\ \emph {et~al.}(2005)\citenamefont {Fang},
  \citenamefont {Wang}, \citenamefont {Zou}, \citenamefont {Tang},
  \citenamefont {Xu}, \citenamefont {Chen}, \citenamefont {Dong}, \citenamefont
  {Shan},\ and\ \citenamefont {Wen}}]{fang_fabrication_2005}%
  \BibitemOpen
  \bibfield  {author} {\bibinfo {author} {\bibfnamefont {L.}~\bibnamefont
  {Fang}}, \bibinfo {author} {\bibfnamefont {Y.}~\bibnamefont {Wang}}, \bibinfo
  {author} {\bibfnamefont {P.~Y.}\ \bibnamefont {Zou}}, \bibinfo {author}
  {\bibfnamefont {L.}~\bibnamefont {Tang}}, \bibinfo {author} {\bibfnamefont
  {Z.}~\bibnamefont {Xu}}, \bibinfo {author} {\bibfnamefont {H.}~\bibnamefont
  {Chen}}, \bibinfo {author} {\bibfnamefont {C.}~\bibnamefont {Dong}}, \bibinfo
  {author} {\bibfnamefont {L.}~\bibnamefont {Shan}}, \ and\ \bibinfo {author}
  {\bibfnamefont {H.~H.}\ \bibnamefont {Wen}},\ }\href {\doibase
  10.1103/PhysRevB.72.014534} {\bibfield  {journal} {\bibinfo  {journal}
  {Physical Review B}\ }\textbf {\bibinfo {volume} {72}},\ \bibinfo {pages}
  {014534} (\bibinfo {year} {2005})}\BibitemShut {NoStop}%
\end{thebibliography}%

\end{document}